\documentclass[conference]{IEEEtran}
\usepackage[
  left=0.673in,
  right=0.673in,
  top=0.747in,
  bottom=0.99in
]{geometry}
\usepackage{multirow}
\usepackage{amsfonts}
\usepackage{placeins}
\usepackage{array}
\usepackage{amsmath,cases,amssymb}
\usepackage{amsmath}
\usepackage{mathtools}
\usepackage{graphicx}
\usepackage{subcaption}
\usepackage{cite}
\usepackage{epsfig}
\usepackage{algorithm}
\usepackage{algorithmic}
\usepackage{epstopdf}
\usepackage{balance}
\usepackage{color}
\usepackage{algorithm}
\DeclareGraphicsExtensions{.eps,.pdf}

\usepackage{scalerel}
\usepackage{multicol}
\usepackage{amssymb}
\usepackage{pifont}
\usepackage{cases}

\usepackage{scalerel}

\let\labelindent\relax
\usepackage{enumitem}

\usepackage{mathbbol}
\usepackage{cite}
\usepackage[T1]{fontenc} 
\usepackage{amsthm} 
\usepackage{eqparbox}
\usepackage{textcomp}
\usepackage[eulergreek]{sansmath}
\usepackage{color}
\usepackage[cmintegrals]{newtxmath}
\usepackage{ctable}
\usepackage{threeparttablex}
\usepackage{makecell}

\newtheorem{remark}{Remark}
\newtheorem{proposition}{Proposition}

\allowdisplaybreaks

\newcommand{\bseq}{\begin{subequations}}
\newcommand{\eseq}{\end{subequations}}
\newcommand{\baln}{\begin{align}}
\newcommand{\ealn}{\end{align}}
\newcommand{\balnd}{\begin{aligned}}
\newcommand{\ealnd}{\end{aligned}}
\newcommand{\beq}{\begin{equation}}
\newcommand{\eeq}{\end{equation}}
\newcommand{\beqn}{\begin{eqnarray}}
\newcommand{\eeqn}{\end{eqnarray}}
\newcommand{\beqno}{\begin{eqnarray*}}
\newcommand{\eeqno}{\end{eqnarray*}}
\newcommand{\bma}{\begin{displaymath}}
\newcommand{\ema}{\end{displaymath}}
\newcommand{\bnu}{\begin{enumerate}}
\newcommand{\enu}{\end{enumerate}}
\newcommand{\bce}{\begin{center}}
\newcommand{\ece}{\end{center}}
\newcommand{\btb}{\begin{tabular}}
\newcommand{\etb}{\end{tabular}}
\newcommand{\bieq}{\begin{IEEEeqnarray}}
\newcommand{\eieq}{\end{IEEEeqnarray}}

\newcommand{\st}{{\mathrm{s.t.}}}
\newcommand{\subnum}{\IEEEyessubnumber}

\makeatletter
\newcommand{\linebreakand}{%
\end{@IEEEauthorhalign}
\hfill\mbox{}\par
\mbox{}\hfill\begin{@IEEEauthorhalign}
}
\makeatother

\begin{document}
\title{DT-Aided Resource Management in Spectrum Sharing Integrated Satellite-Terrestrial Networks }
\vspace{-3mm}

\author{\IEEEauthorblockN{Hung Nguyen-Kha${}^{\dagger}$, Vu Nguyen Ha${}^{\dagger}$, Ti Ti Nguyen${}^{\dagger}$, Eva Lagunas${}^{\dagger}$, Symeon Chatzinotas${}^{\dagger}$, and Joel Grotz${}^{\ddagger}$}

\IEEEauthorblockA{\textit{${}^{\dagger}$Interdisciplinary Centre for Security, Reliability and Trust (SnT), University of Luxembourg, Luxembourg} \\
       \textit{${}^{\ddagger}$SES S.A., Luxembourg}}
       
       \vspace{-10mm}
       }


%

\maketitle

\begin{abstract} 
The integrated satellite-terrestrial networks (ISTNs) through spectrum sharing have emerged as a promising solution to improve spectral efficiency and meet increasing wireless demand. However, this coexistence introduces significant challenges, including inter-system interference (ISI) and the low Earth orbit satellite (LSat) movements. To capture the actual environment for resource management, we propose a time-varying digital twin (DT)-aided framework for ISTNs incorporating 3D map that enables joint optimization of bandwidth (BW) allocation, traffic steering, and resource allocation, and aims to minimize congestion. The problem is formulated as a mixed-integer nonlinear programming (MINLP), addressed through a two-phase algorithm based on successive convex approximation (SCA) and compressed sensing approaches. Numerical results demonstrate the proposed method’s superior performance in queue length minimization compared to benchmarks.
\end{abstract}

\vspace{-0.25 cm}
\section{Introduction}
\vspace{-2mm}
Recently, sharing the same radio frequency band (RFB) between terrestrial network (TN) and non-terrestrial networks (NTN) systems has become a promising solution to enhance spectral efficiency and support the growing demand for wireless connectivity
\cite{FCC2322_Supplemental_Space_Coverage}. 
However, sharing the same RFBs introduces the critical ISI issue. Additionally, the movements of users (UEs) and  LSat further introduce dynamics to the systems. Hence, optimizing systems to manage the ISI issue and terminal movement poses significant challenges, especially in complex environments such as urban areas. 

The spectrum sharing ISTNs has been studied in the literature such as \cite{okati_arXiv2401.08453_Sband, kim_spectrum_2024, Lee_TVT23, Li_TWC2024, martikainen_WoWMoM2023, Zhu_TWC24_BeamManage}. Particularly, in \cite{okati_arXiv2401.08453_Sband, kim_spectrum_2024}, the performance analysis is investigated by using the stochastic geometry approach. In \cite{Lee_TVT23, Li_TWC2024}, the snapshot resource management in different scenarios are studied. However, the consideration of the snapshot model in these works struggles to capture the dynamic systems in ISTNs.
Subsequently, the spectrum sharing problem is studied in \cite{martikainen_WoWMoM2023, Zhu_TWC24_BeamManage} limited to the system level. Almost all existing works rely on the statistical channel model, which is challenging to capture actual environments.


In this work, we study the time-varying DT-aided ISTNs where TN and NTN systems share the same RFB. We aim to optimize BW allocation, traffic steering, UE and resource block (RB) assignment, and power control under the service constraint to minimize the congestion, i.e., queue length minimization. The underlying MINLP  problem is challenging to solve. Additionally, the problem requires information of channel gain and arrival traffic of the upcoming time cycle (TC)
which further poses challenge in solving. 
To assist resource management, we utilize a DT model with the 3D map to capture the system. Furthermore, we propose a two-phase algorithm based on the SCA technique. First, the problem for a TC is solved based on the predicted information from the DT model. Subsequently, the power control is re-optimized based on the instant estimated channel and initial point provided by the first algorithm's outcome. The greedy algorithm is further proposed for comparison purposes. The numerical result shows the effectiveness in terms of queue length minimization. 
\begin{figure}
	\centering
	\includegraphics[width=8.5cm, height=2.6cm]{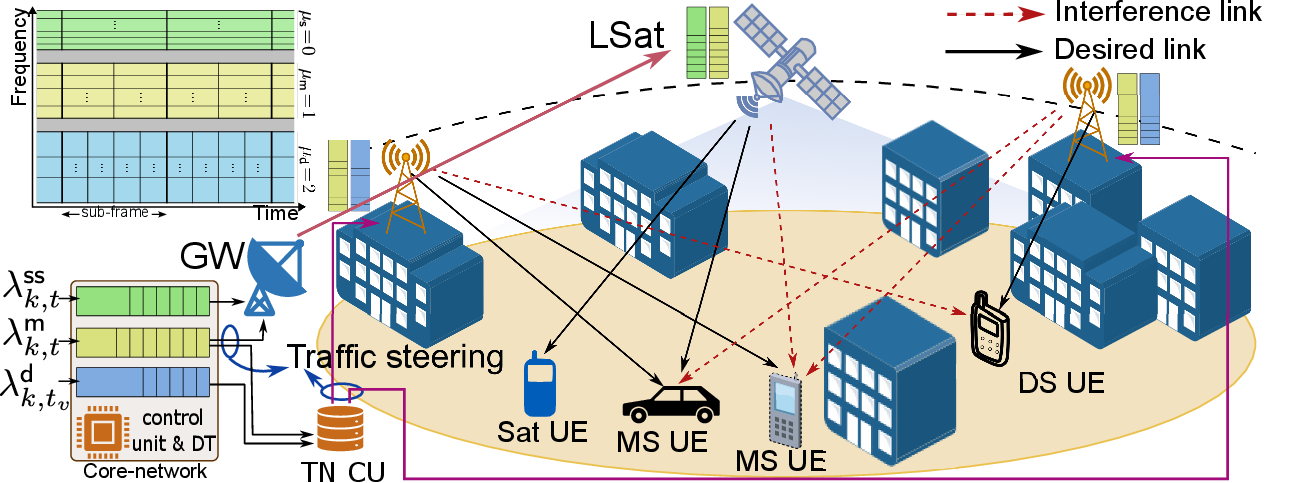}
    \vspace{-2mm}
	\caption{System model.}
	\label{fig:sysmodel}
    \vspace{-4mm}
\end{figure}

\vspace{-0.3 cm}
\section{System Model}
\vspace{-2mm}
\enlargethispage{-0.05 cm} 
We consider the downlink (DL) ISTNs including $N$ access points (APs), one LSat, and $K$ users described in Fig.~\ref{fig:sysmodel}.
Let $\mathcal{N} = \{1,\dots, N \}$ and $\mathcal{U} = \{1,\dots, K \}$, ${\sf{AP}}_{n}$, and ${\sf{UE}}_{k}$ denote the sets of APs and UEs, the $n-$th AP and $k-$th UE, where index $n=0$ indicates the LSat, respectively. 
The user set is divided into three sets using delay-sensitive (DS), satellite (SS) and mobile (MS) services denoted by $\mathcal{U}^{\sf{d}}$, $\mathcal{U}^{\sf{s}}$, and $\mathcal{U}^{\sf{m}}$ served by APs, LSat, and both, respectively.
For convenience, let $\mathcal{S}^{\sf{tn}} = \{\sf{d,m}\}$, $\mathcal{S}^{\sf{sn}} = \{\sf{m,s}\}$, and $\mathcal{S} = \{\sf{d,s,m}\}$ be the service sets. 
We consider systems in $N_{\sf{cy}}$ TCs, each TC indexed by $c$ consists of $N^{\sf{tf}}$ time-frame (TF). 
We assume that the ISTNs operate in the shared RFB with the BW of $B$ which can be dynamically allocated to services, TNs, and satellite networks (SNs) over TCs. 
Particularly, the entire BW is divided into three bandwidth parts (BWPs) used for DS, MS, and SS services, respectively. 
Furthermore, we assume that the OFDMA and 5G NR frameworks are utilized where numerologies $\mu_{\sf{x}}$ are used for BWPs $\sf{x} \in \mathcal{S}$ with $\mu_{\sf{s}} = 0, \mu_{\sf{m}} = 1$ and $\mu_{\sf{d}} = 2$ \cite{Kihero_Access19}, respectively.
Each TF consists of $10$ sub-frames (SFs). Assuming that TNs and SNs operate in TDD and FDD modes, TNs use $N_{\sf{sf}}^{\sf{tn,dl}}$ beginning SFs while SNs use $10$ SFs for DL in each TF. 
The radio resource is divided into a grid of RBs where the RB size (duration and BW) is scaled according to numerology $\mu$ with the parameters described in Table.~\ref{tab:RBGrid}. 
To avoid the inter-numerology-interference, we assume that there are guard bands with fixed BWs of $B_{\sf{G}}^{\sf{sm}}=w_{\sf{s}}$ and $B_{\sf{G}}^{\sf{md}}=w_{\sf{m}}$ between BWPs $\sf{s}-\sf{m}$ and $\sf{m}-\sf{d}$, respectively. 
To describe the BW allocation, we introduce the binary variable $\boldsymbol{b}_{c} \triangleq \{b_{f^{\sf{x}},c}^{\sf{x}} | \forall f^{\sf{x}} \in \mathcal{F}^{\sf{x}}, {\sf{x}} \in \mathcal{S} \}$ where $b_{f^{\sf{x}},c}^{\sf{x}} = 1$ if sub-band (SB) $f^{\sf{x}}$ service $\sf{x}$ used at TC $c$ and $=0$ in otherwise.
To ensure the non-overlap between BWPs, the continuity in BWPs, and maximum BW, one needs to hold constraints
\bieq{ll}
    (C1): \; b_{f^{\sf{d}},c}^{\sf{d}} + \scaleobj{0.8}{\sum\nolimits_{i=1}^{\bar{Q}^{\sf{d}}_{\sf{m}} (f^{\sf{d}}+0.5) }} b_{i,c}^{\sf{m}} + \scaleobj{0.8}{\sum\nolimits_{j=1}^{\bar{Q}^{\sf{d}}_{\sf{s}} (f^{\sf{d}}+0.5) }} b_{j,c}^{\sf{s}} \leq 1, \; \forall f^{\sf{d}}, c, \nonumber \\
    (C2): \; b_{f^{\sf{m}},c}^{\sf{m}} +  \scaleobj{0.8}{\sum\nolimits_{i=1}^{\bar{Q}^{\sf{m}}_{\sf{s}} (f^{\sf{m}}+0.5) }  } b_{i,c}^{\sf{s}} \leq 1, \; \forall f^{\sf{m}}, \forall c, \nonumber \\
    (C3): \; \scaleobj{0.8}{\sum\nolimits_{\sf{x} \in \{d,m,s\}}} w_{\sf{x}} \scaleobj{0.8}{\sum\nolimits_{\forall f^{\sf{x}} \in \mathcal{F}^{\sf{x}}}} b_{f^{\sf{x}},c}^{\sf{x}} + B_{\sf{G}}^{\sf{sm}} + B_{\sf{G}}^{\sf{md}} \leq B, \; \forall c. \nonumber
\eieq
\vspace{-4mm}

\vspace{-2mm}
\subsection{Digital-Twin Model}
\vspace{-1mm}
The DT model includes the replication of features and relevant information of UEs, APs, LSats, and environments.
In cell $n$, the DT model of environment in its area and ${\sf{AP}}_{n}$ are
$
{\sf{DT}}_{n}^{\sf{env}} = \{{\sf{map}}_{n}\}, \;
{\sf{DT}}_{n}^{\sf{ap}} = \{ \boldsymbol{u}_{n}^{\sf{ap}} \}, \; \forall n \in \mathcal{N}, 
$
where ${\sf{map}}_{n}$ and $\boldsymbol{u}_{n}^{\sf{ap}}$ are 3D map of cell $n$ area ${\sf{AP}}_{n}$'s position \cite{Hung_VTC24}.
The DTs of $\sf{UE}_{k}$ and the LSat at TF $t$ are modeled as 
$
{\sf{DT}}_{k,t}^{\sf{ue}} = \{ \hat{\boldsymbol{u}}_{k,t}^{\sf{ue}} , \hat{\lambda}_{k,t}^{\sf{x}} \},  \forall k \in \mathcal{U}^{\sf{x}}, {\sf{x}}\in \mathcal{S}, \;{\sf{DT}}_{t}^{\sf{sat}} = \{ \boldsymbol{u}_{t}^{\sf{sat}} \},
$ where 
$\hat{\boldsymbol{u}}_{k,t}^{\sf{ue}}$, $\boldsymbol{u}_{t}^{\sf{sat}}$ and $\hat{\lambda}_{k,t}^{\sf{x}}$ are the ${\sf{UE}}_{k}$, LSat positions and ${\sf{UE}}_{k}$ arrival rate in the DT model obtained by the updated real information.

\vspace{-2mm}
\enlargethispage{-0.05 cm} 
\subsection{Channel Model}
Let  $\tilde{h}_{n,k}^{f^{\sf{x}},t_{s}^{\sf{x}}}$ be the channel coefficient between ${\sf{AP}}_{n}$ and ${\sf{UE}}_{k}$ over SB $f^{\sf{x}}$ in BWP $\sf{x} \in \mathcal{S}^{\sf{tn}}$ at TS $t_{s}^{\sf{x}}$. 
Regarding the LSat-UE links, let $\tilde{g}_{m,k}^{f^{\sf{x}},t_{s}^{\sf{x}}}$ be the channel coefficient between ${\sf{LSat}}_{m}$ and ${\sf{UE}}_{k}$ over SB $f^{\sf{x}}$ in BWP  $\sf{x \in \mathcal{S}^{\sf{sn}}}$ at TS $t_{s}^{\sf{x}}$.
The channels of AP-UE and LSat-UE links are modeled based on the Rician channel model and ray-tracing (RT) mechanism. For simplicity, let's us omit the index ${\sf{x}}$ in the RB's index in this section, channels coefficients $\tilde{h}_{n,k}^{f,t_{s}}$ and $\tilde{g}_{m,k}^{f,t_{s}}$ are modeled as
$
    \tilde{h}_{n,k}^{f,t_{s}} = \scaleobj{0.8}{\sqrt{{\sf{PL}}_{n,k}^{f,t_{s}}}} \scaleobj{0.8}{( \sqrt{\frac{\tilde{K}}{\tilde{K}+1}}} \tilde{h}_{n,k}^{{\sf{los}},f,t_{s}} + \scaleobj{0.8}{\sqrt{\frac{1}{\tilde{K}+1}}} \tilde{h}_{n,k}^{{\sf{nlos}},f,t_{s}}  ), \;   
	\tilde{g}_{k}^{f,t_{s}} = \scaleobj{0.8}{\sqrt{{\sf{PL}}_{k}^{f,t_{s}}}} ( \scaleobj{0.8}{\sqrt{\frac{\tilde{K}}{\tilde{K}+1}}} \tilde{g}_{k}^{{\sf{los}},f,t_{s}} + \scaleobj{0.8}{\sqrt{\frac{1}{\tilde{K}+1}}} \tilde{g}_{k}^{{\sf{nlos}},f,t_{s}}  ),
$
wherein $\tilde{K}$ is the K-factor, ${\sf{PL}}_{n,k}^{f,t_{s}}$, $\tilde{h}_{n,k}^{{\sf{los}},f,t_{s}}$, $\tilde{h}_{n,k}^{{\sf{nlos}},f,t_{s}}$, ${\sf{PL}}_{k}^{f,t_{s}}$, $\tilde{g}_{k}^{{\sf{los}},f,t_{s}}$, and $\tilde{g}_{k}^{{\sf{nlos}},f,t_{s}}$ indicate the path-loss, line-of-sight (LoS) and non-line-of-sight (NLoS) components of ${\sf{AP}}_{n} - {\sf{UE}}_{k}$ and ${\sf{LSat}} - {\sf{UE}}_{k}$ links at the corresponding frequency and time, respectively. 

To reflect the environment features in considered area, the RT mechanism is utilized with APs, LSat, and UEs positions and 3D map of considered areas in generating the channel \cite{Hung_VTC24}. 
Since the 3D map data is incomplete for all objects in the environment, the RT mechanism cannot compute the full channel paths. Hence, the NLoS components are further modeled as
$
    \tilde{h}_{n,k}^{{\sf{nlos}},f,t_{s}} = \sqrt{\xi} \bar{h}_{n,k}^{{\sf{nlos}},f,t_{s}} + \sqrt{(1 - \xi)} \delta_{n,k}^{f,t_{s}}$ and $ 
    \tilde{g}_{k}^{{\sf{nlos}},f,t_{s}} = \sqrt{\xi} \bar{g}_{k}^{{\sf{nlos}},f,t_{s}} + \sqrt{(1 - \xi)} \delta_{k}^{f,t_{s}}, 
$
where $\bar{h}_{n,k}^{{\sf{nlos}},f,t_{s}}$ and $\bar{g}_{k}^{{\sf{nlos}},f,t_{s}}$ indicate the NLoS components identified by the RT mechanism while $\delta_{n,k}^{f,t_{s}}$ and $\delta_{k}^{f,t_{s}}$ indicate the NLoS components resulting from objects absent  in the 3D map data. Hence, they can be modeled as complex normal random variables with zero mean and unit variance. $\xi \in (0,1)$ is the factor wherein $\xi = 1$ indicates that the NLoS component is deterministic while $\xi = 0$ means 
the NLoS component follows the traditional Rician model.
\begin{table}[h]
\footnotesize
\captionsetup{font=small}
\centering
\scalebox{0.7}{
\begin{tabular}{|l|l|l|l|}
\hline
                &    Index  & Duration (ms) \\ \hline 
     Frame (TF)  &     $t \in \mathcal{T}_{c}^{\sf{tf}} 
        \triangleq \{(c-1)N_{\sf{tf}}+1,\dots,cN_{\sf{tf}}\}$ & $T_{\sf{f}}=10$ \\
     Sub-frame (SF) &  $t_{v} 
        \in \mathcal{T}_{t}^{\sf{sf}} \triangleq \{ 10(t-1)+v, v \in \{1,\dots,10\} \}$  & $T_{\sf{sf}}=1$  \\
     \makecell[l]{Time-slot (TS) \\ service $\sf{x}$} &  \makecell{$t_s^{\sf{x}} \in 
        \mathcal{T}_{t_{\sf{v}}}^{\sf{sf,x}} \triangleq \{ (t_{v}-1) N_{\sf{ts}}^{\sf{sf,x}} + 1,\dots, t_{v} N_{\sf{ts}}^{\sf{sf,x}}\} $ \\
        $\subset \mathcal{T}_{t}^{\sf{x}} \triangleq \{ (t-1) N_{\sf{ts}}^{\sf{x}} + 1,\dots, t N_{\sf{ts}}^{\sf{x}} \}$}   & $T_{\sf{x}}=2^{-\mu_{\sf{x}}}$  \\ \hline
     \multicolumn{3}{|l|}{$N_{\sf{tf}}$: \#TF/cycle,  $N_{\sf{ts}}^{\sf{x}}$: \#TS/TF, $N_{\sf{ts}}^{\sf{sf,x}}$: \#TS/SF, $N_{\sf{sf}} = 10$ SFs/TF} \\ 
     \multicolumn{3}{|l|}{$\mathcal{T}_{c}^{\sf{tf}} = \{ (c-1) N_{\sf{tf}}+1, \dots, c N_{\sf{tf}} \}$: TF set in TC $c$ } \\
     \multicolumn{3}{|l|}{$\mathcal{T}_{c}^{\sf{ts,x}} = \{t_{s}^{\sf{x}} | \forall t_{s}^{\sf{x}} \in \mathcal{T}_{t}^{\sf{x}}, \forall t \in \mathcal{T}_{c}^{\sf{tf}} \}$: TS set in TC $c$ } \\
     \hline
                & Index & Bandwidth (kHz)\\ \hline
     \makecell[l]{Sub-band (SB) \\ service $\sf{x}$} 
        & $f^{\sf{x}} \in \mathcal{F}^{\sf{x}} \triangleq \{ 1,\dots, F^{\sf{x}} \}$ 
        & $w_{\sf{x}} = 2^{\mu_{\sf{x}}} \times 180$ \\ \hline
    \multicolumn{3}{|l|}{ $F^{\sf{x}}=\lfloor B/w_{\sf{x}}\rfloor$: Num. SBs service $\sf{x}$, $\bar{Q}^{\sf{x}}_{\sf{x'}}=2^{\mu_{\sf{x}} - \mu_{\sf{x'}}}$: RB BW ratio,  } \\ \hline
\end{tabular}}
\vspace{-1mm}
\caption{Parameter of the resource block grid.}
\label{tab:RBGrid}
\vspace{-3mm}
\end{table}

\vspace{-2mm}
\enlargethispage{-0.05 cm} 
\subsection{Transmission Model}
\vspace{-2mm}
\subsubsection{Association Model}
Let $\boldsymbol{\alpha} = [\alpha_{n,k}^{f^{\sf{x}},t_{s}^{\sf{x}}}]$ 
where $\alpha_{n,k}^{f^{\sf{x}},t_{s}^{\sf{x}}} =1$ if ${\sf{UE}}_{k}$ served by ${\sf{AP}}_{n}$ over ${\sf{RB}}(f^{\sf{x}}, t_{s}^{\sf{x}})$ and $s \leq N_{\sf{sf}}^{\sf{tn,dl}} N_{\sf{ts}}^{\sf{sf},\sf{x}}$, and $=0$ otherwise. It is aligned with BW allocation as
\vspace{-2mm}
\beq
    (C4)\!:  \alpha_{n,k}^{f^{\sf{x}},t_{s}^{\sf{x}}} \! \leq \! b_{f^{\sf{x}},c}^{\sf{x}}, \forall (k,\!f^{\sf{x}},\!t_{x}^{\sf{s}}) \!\in \! \mathcal{U}^{\sf{x}}  \times \mathcal{F}^{\sf{x}} \times \mathcal{T}_{c}^{\sf{ts,x}}, {\sf{x}} \!\in\! \mathcal{S}^{\sf{tn}}, \! \forall n, c. \nonumber
\eeq
To ensure the orthogonality at each AP, one assumes that each RB of each AP can be assigned to at most one UE, as follows
\beq
    (C5)\!: \scaleobj{0.87}{\sum\nolimits_{\forall k}} \alpha_{n,k}^{f^{\sf{x}},t_{s}^{\sf{x}}} \leq 1, \;  \forall (f^{\sf{x}}, t_{s}^{\sf{x}}) \in \mathcal{F}^{\sf{x}} \times \mathcal{T}_{c}^{\sf{ts,x}}, {\sf{x}} \in \mathcal{S}^{\sf{tn}}, \forall n,c. \nonumber
\eeq
Regarding multi-connectivity, each $\sf{UE}^{\sf{d}}$ can be served by multiple APs at each TS via different RBs, as follows
\beq
    (C6)\!:  \scaleobj{0.8}{\sum\nolimits_{\forall n}} \alpha_{n,k}^{f^{\sf{d}},t_{s}^{\sf{d}}} \leq 1, \; \forall k \in \mathcal{U}^{\sf{d}}, \forall (f^{\sf{d}}, t_{s}^{\sf{d}}) \in \mathcal{F}^{\sf{d}} \times \mathcal{T}_{c}^{\sf{ts,d}}, \forall c. \nonumber
\eeq
Regarding LSat-UE association, we introduce a variable $\boldsymbol{\beta}= [\beta_{k}^{f^{\sf{x}},t_{s}^{\sf{x}}}]$ 
where $\beta_{k}^{f^{\sf{x}},t_{s}^{\sf{x}}} = 1$ if ${\sf{UE}}_{k}$ served by LSat over ${\sf{RB}}(f^{\sf{x}}, t_{s}^{\sf{x}})$, and $=0$ otherwise.
Similar to the AP-UE association, the LSat-UE association must satisfy
\beq
    (C7)\!: \; \beta_{k}^{f^{\sf{x}},t_{s}^{\sf{x}}} \!\leq \!b_{f^{\sf{x}},c}^{\sf{x}}, \forall (k, f^{\sf{x}}, t_{x}^{\sf{x}}) \!\in\! \mathcal{U}^{\sf{x}} \times \mathcal{F}^{\sf{x}} \times \mathcal{T}_{c}^{\sf{ts,x}}, {\sf{x}} \in \mathcal{S}^{\sf{sn}}, \forall c. \nonumber
\eeq
The orthogonality between UEs served by the LSat is 
\bieq{ll}
    (C8)\!:  \scaleobj{0.8}{\sum\nolimits_{\forall k}} \beta_{k}^{f^{\sf{x}},t_{s}^{\sf{x}}} \leq 1, \; \forall (f^{\sf{x}}, t_{s}^{\sf{x}}) \in \mathcal{F}^{\sf{x}} \times \mathcal{T}_{c}^{\sf{ts,x}}, {\sf{x}} \in \mathcal{S}^{\sf{sn}}, \forall c. \nonumber
\eieq
Additionally, we assume that UEs using MS services can be served by both the BSs and LSat at the same time via different RBs in BWP $\sf{m}$, which yields the constraint
\beq
    (C9)\!: \; \scaleobj{0.8}{\sum\nolimits_{\forall n}} \alpha_{n,k}^{f^{\sf{m}},t_{s}^{\sf{m}}} + \beta_{k}^{f^{\sf{m}},t_{s}^{\sf{m}}} \leq 1, \; \forall k \in \mathcal{U}^{m}, \forall (f^{\sf{m}},t_{s}^{\sf{m}}), \forall c. \nonumber
\eeq

\enlargethispage{-0.05 cm} 
\subsubsection{Transmission over BWP ${\sf{d}}$}
This BWP is utilized only by APs to serve DS UEs. The ${\sf{UE}}_{k}^{\sf{d}}$ received signal over ${\sf{RB}}(f^{\sf{d}},t_{s}^{\sf{d}})$ is expressed as
$
    y_{k}^{f^{\sf{d}},t_{s}^{\sf{d}}} = f_{k}^{\sf{rx,tn,d}}(\boldsymbol{p}_{t}, \boldsymbol{\alpha}_{t}, \boldsymbol{x}_{t}) + \varsigma_{k}^{f^{\sf{d}},t_{s}^{\sf{d}}} ,
$
where $f_{k}^{\sf{rx,tn,x}}(\boldsymbol{p}_{t}, \boldsymbol{\alpha}_{t}, \boldsymbol{x}_{t}) = {\sum_{\forall i} \sum_{\forall j }} \scaleobj{0.8}{\sqrt{ \alpha_{i,j}^{f^{\sf{x}},t_{s}^{\sf{x}}} p_{i,j}^{f^{\sf{x}}, t_{s}^{\sf{x}}} }} \tilde{h}_{i,k}^{f^{\sf{x}},t_{s}^{\sf{x}}} x_{i,j}^{f^{\sf{x}}, t_{s}^{\sf{x}}}$, $p_{n,k}^{f^{\sf{d}},t_{s}^{\sf{d}}}$ and $x_{n,k}^{f^{\sf{d}},t_{s}^{\sf{d}}}$ are the transmission power and symbol from ${\sf{AP}}_{n}$ to ${\sf{UE}}_{k}$ over ${\sf{RB}}(f^{\sf{d}},t_{s}^{\sf{d}})$. $\varsigma_{k}^{f^{\sf{d}},t_{s}^{\sf{d}}}$ is the additive Gaussian noise at ${\sf{UE}}_{k}$.
Hence, the corresponding signal-to-interference-plus-noise-ratio (SINR) is expressed as
\vspace{-2mm}
\beq\label{eq: SINR BWPds}
    \gamma_{n,k}^{f^{\sf{d}},t_{s}^{\sf{d}}}(\boldsymbol{p}_{t}, \boldsymbol{\alpha}_{t}) = \scaleobj{0.8}{{\alpha_{n,k}^{f^{\sf{d}},t_{s}^{\sf{d}}} p_{n,k}^{f^{\sf{d}},t_{s}^{\sf{d}}} h_{n,k}^{f^{\sf{d}},t_{s}^{\sf{d}}}}/{ \big( \Psi_{n,k}^{f^{\sf{d}},t_{s}^{\sf{d}}}(\boldsymbol{p}_{t}, \boldsymbol{\alpha}_{t}) + \sigma_{{\sf{d}},k}^2 \big)}},
\eeq
\vspace{-2mm}
with $\boldsymbol{p}_{t} \triangleq \{ p_{n,k}^{f^{\sf{x}},t_{s}^{\sf{x}}} | {\forall n, \forall f_n^{\sf{x}}, \forall k \in \mathcal{U}^{\sf{x}}, \forall t_{s}^{\sf{x}} \in \mathcal{T}_{t}^{\sf{x}} , {\sf{x \in \{ 
d,m \}}}} \}$, $\boldsymbol{\alpha}_{t} \triangleq \{ \alpha_{n,k}^{f^{\sf{x}},t_{s}^{\sf{x}}} | {\forall n, \forall f_n^{\sf{x}}, \forall k \in \mathcal{U}^{\sf{x}}, \forall t_{s}^{\sf{x}} \in \mathcal{T}_{t}^{\sf{x}} , {\sf{x}} \in \mathcal{S}^{\sf{tn}} } \}$, and $h_{n,k}^{f^{\sf{x}},t_{s}^{\sf{x}}} = |\tilde{h}_{n,k}^{f^{\sf{x}},t_{s}^{\sf{x}}}|^2$. $\Psi_{n,k}^{f^{\sf{x}},t_{s}^{\sf{x}}}(\boldsymbol{p}_{t}, \boldsymbol{\alpha}_{t}) = {\sum_{\forall i \neq n} \sum_{\forall j}} \alpha_{i,j}^{f^{\sf{x}},t_{s}^{\sf{x}}} p_{i,j}^{f^{\sf{x}}, t_{s}^{\sf{x}}} h_{i,k}^{f^{\sf{x}},t_{s}^{\sf{x}}}$ is the ICI power at ${\sf{UE}}_{k}^{\sf{x}}$. 
Due to the delay requirement of the DS services, we assume that DS data is transmitted within each SF and the short packet framework is used for the transmission of DS services. Hence,  according to \cite{polyanskiy_FiniteCode}, the aggregated achievable rate of ${\sf{UE}}_{k}^{\sf{d}}$ served by ${\sf{AP}}_{n}$ at SF $t_{v}^{\sf{d}}$ can be expressed as 
$
    R_{n,k}^{{\sf{d}}, t_{v}^{\sf{d}}}(\boldsymbol{p}_{t},\boldsymbol{\alpha}_{t}) = w_{\sf{d}} \sum_{\forall f^{\sf{d}} \in \mathcal{F}^{\sf{d}}} \sum_{\forall t_{s}^{\sf{d}} \in \mathcal{T}_{t_{v}}^{\sf{sf,d}}} \Big( \ln{(1 + \gamma_{n,k}^{f^{\sf{d}},t_{s}^{\sf{d}}}(\boldsymbol{p}_{t}, \boldsymbol{\alpha}_{t})}) \nonumber \\
     - \scaleobj{0.8}{{\alpha_{n,k}^{f^{\sf{d}},t_{s}^{\sf{d}}} \sqrt{V_{n,k}^{f^{\sf{d}}, t_{s}^{\sf{d}}}} Q^{-1}(P_{\epsilon}) }/{\sqrt{\sum_{\forall f^{\sf{d}} \in \mathcal{F}^{\sf{d}}} \sum_{\forall t_{s}^{\sf{d}} \in \mathcal{T}_{t_{v}}^{\sf{sf,d}}} \alpha_{n,k}^{f^{\sf{d}},t_{s}^{\sf{d}}} \tau_{\sf{d}} w_{\sf{d}}} } } \Big),
$
wherein $V_{n,k}^{f^{\sf{d}}, t_{s}^{\sf{d}}} = 1 - (1 + \gamma_{n,k}^{f^{\sf{d}},t_{s}^{\sf{d}}}(\boldsymbol{p}_{t}, \boldsymbol{\alpha}_{t}))^{-2}, \; Q^{-1}(\cdot)$ and $P_{\epsilon}$ are the channel dispersion, the inverse of the Q-function, and the error probability. The channel dispersion can be approximated as $V_{n,k}^{f^{\sf{d}}, t_{s}^{\sf{d}}} \approx 1$ for a sufficiently high $\gamma_{n,k}^{f^{\sf{d}},t_{s}^{\sf{d}}}(\boldsymbol{p}^{\sf{a}}, \boldsymbol{\alpha}_{t}) \geq \gamma_{0}^{\sf{d}}$ with $\gamma_{0}^{\sf{d}} \geq 5$~dB \cite{schiessl_Fading_FiniteCode}. To guarantee the channel dispersion approximation, we consider constraint
\vspace{-2mm}
\beq
    (C10): \quad \gamma_{n,k}^{f^{\sf{d}},t_{s}^{\sf{d}}}(\boldsymbol{p}_{t}, \boldsymbol{\alpha}_{t}) \geq \alpha_{n,k}^{f^{\sf{d}},t_{s}^{\sf{d}}} \gamma_{0}^{\sf{d}}, \; \forall (n,k) \in \mathcal{N} \times \mathcal{U}^{\sf{d}}. \nonumber
\eeq
Subsequently, let  $\chi_{\sf{d}} = \sqrt{w_{\sf{d}}}\sqrt{V} Q^{-1}(P_{\epsilon})/\sqrt{\tau_{\sf{d}} }$ and $V \approx V_{n,k}^{f^{\sf{d}}, t_{s}^{\sf{d}}} \approx 1$, the achievable rate 
$
    R_{n,k}^{{\sf{d}}, t_{v}^{\sf{d}}}(\boldsymbol{p}_{t},\boldsymbol{\alpha}_{t}) $ $= w_{\sf{d}}\!\!\!\!\! \sum\limits_{\forall f^{\sf{d}} \in \mathcal{F}^{\sf{d}}} \sum\limits_{\forall t_{s}^{\sf{d}} \in \mathcal{T}_{t_{v}}^{\sf{sf,d}}}  \ln{(1 + \gamma_{n,k}^{f^{\sf{d}},t_{s}^{\sf{d}}}(\boldsymbol{p}_{t}, \boldsymbol{\alpha}_{t})})  {-} \chi_{\sf{d}} \scaleobj{0.8}{\sqrt{\sum\limits_{\forall f^{\sf{d}} \in \mathcal{F}^{\sf{d}}} \sum\limits_{\forall t_{s}^{\sf{d}} \in \mathcal{T}_{t_{v}}^{\sf{sf,d}}} \alpha_{n,k}^{f^{\sf{d}},t_{s}^{\sf{d}}} }  } .
    $


\subsubsection{Transmission over BWP $\sf{m}$}
In this BWP, UEs can be served by APs and the LSat. 
First, the received signal at ${\sf{UE}}_{k}^{\sf{m}}$ over ${\sf{RB}}(f^{\sf{m}},t_{s}^{\sf{m}})$ of ${\sf{BWP}}_{t}^{\sf{m}}$ can be expressed as
\bieq{ll}
    y_{k}^{f^{\sf{m}},t_{s}^{\sf{m}}} = f_{k}^{\sf{rx,tn,m}}(\boldsymbol{p}_{t}, \boldsymbol{\alpha}_{t}, \boldsymbol{x}_{t}) + \varsigma_{k}^{f^{\sf{m}},t_{s}^{\sf{m}}} + f_{k}^{\sf{rx,sn,m}}(\boldsymbol{P}_{t}, \boldsymbol{\beta}_{t}, \boldsymbol{x}_{t}). \nonumber
\eieq
with $f_{k}^{\sf{rx,sn,x}}(\boldsymbol{P}_{t}, \boldsymbol{\beta}_{t}, \boldsymbol{x}_{t}) = \sum_{\forall j} \scaleobj{0.8}{\sqrt{\beta_{j}^{f^{\sf{m}},t_{s}^{\sf{m}}} P_{k}^{f^{\sf{m}}, t_{s}^{\sf{m}}} }} \tilde{g}_{k}^{f^{\sf{m}},t_{s}^{\sf{m}}} x_{j}^{f^{\sf{m}}, t_{s}^{\sf{m}}}$. 
Assuming that ${\sf{UE}}_{k}^{\sf{m}}$ is served by ${\sf{AP}}_{n}$, the SINR is written as
\beq\label{eq: SINR BWPms}
    \gamma_{n,k}^{f^{\sf{m}},t_{s}^{\sf{m}}}(\boldsymbol{p}_{t}, \boldsymbol{P}_{t}, \boldsymbol{\alpha}_{t}, \boldsymbol{\beta}_{t}) = \scaleobj{0.8}{\frac{\alpha_{n,k}^{f^{\sf{m}},t_{s}^{\sf{m}}} p_{n,k}^{f^{\sf{m}},t_{s}^{\sf{m}}} h_{n,k}^{f^{\sf{m}},t_{s}^{\sf{m}}}}{ \Psi_{n,k}^{f^{\sf{m}},t_{s}^{\sf{m}}}(\boldsymbol{p}_{t}, \boldsymbol{\alpha}_{t}) + \Theta_{k}^{{\sf{a}},f^{\sf{m}},t_{s}^{\sf{m}}}(\boldsymbol{P}_{t}, \boldsymbol{\beta}_{t}) + \sigma_{{\sf{m}},k}^2}}, \nonumber
\eeq
and $\Theta_{k}^{{\sf{a}},f^{\sf{m}},t_{s}^{\sf{m}}}(\boldsymbol{P}_{t}, \boldsymbol{\beta}_{t}) =  \scaleobj{0.8}{\sum_{\forall k'}} \beta_{k'}^{f^{\sf{m}},t_{s}^{\sf{m}}} P_{k'}^{f^{\sf{m}}, t_{s}^{\sf{m}}}  g_{k}^{f^{\sf{m}},t_{s}^{\sf{m}}}$ is the ISI from the LSat, 
where $P_{k}^{f^{\sf{x}},t_{s}^{\sf{x}}}$ is the transmit power from the LSat to ${\sf{UE}}_{k}$ over ${\sf{RB}}(f^{\sf{x}},t_{s}^{\sf{x}})$ of BWP $\sf{x}$, $\boldsymbol{P}_{t} \triangleq \{ P_{k}^{f^{\sf{x}},t_{s}^{\sf{x}}} | { \forall f^{\sf{x}},\forall k \in \mathcal{U}^{\sf{x}}, \forall t_{s}^{\sf{x}} \in \mathcal{T}_{t}^{\sf{x}}, {\sf{x \in \{m,s\}}}} \}$, $\boldsymbol{\beta}_{t} \triangleq \{ \beta_{k}^{f^{\sf{x}},t_{s}^{\sf{x}}} | {\forall f^{\sf{x}},\forall k \in \mathcal{U}^{\sf{x}}, \forall t_{s}^{\sf{x}} \in \mathcal{T}_{t}^{\sf{x}}, {\sf{x \in \{m,s\}}}} \}$, and $g_{k}^{f^{\sf{m}},t_{s}^{\sf{m}}} = |\tilde{g}_{k}^{f^{\sf{m}},t_{s}^{\sf{m}}}|^2$. 
The achievable rate from ${\sf{AP}}_{n}$ to ${\sf{UE}}_{k}^{\sf{m}}$ over ${\sf{RB}}(f^{\sf{m}},t_{s}^{\sf{m}})$ and TS $t_{s}^{\sf{m}}$ is
\bieq{ll} \label{eq: rate RB ms ap}
    R_{n,k}^{{\sf{m}},f^{\sf{m}},t_{s}^{\sf{m}}}\!\!(\boldsymbol{p}_{t},\! \boldsymbol{P}_{t}, \!\boldsymbol{\alpha}_{t},\! \boldsymbol{\beta}_{t}) {=} w_{\sf{m}} \! \ln \!\left( 1 + \gamma_{n,k}^{f^{\sf{m}}, t_{s}^{\sf{m}}}(\boldsymbol{p}_{t}, \!\boldsymbol{P}_{t}, \!\boldsymbol{\alpha}_{t}, \!\boldsymbol{\beta}_{t}) \! \right)\!, \subnum \\
    R_{n,k}^{{\sf{m}},t_{s}^{\sf{m}}}(\boldsymbol{p}_{t},\! \boldsymbol{P}_{t},\! \boldsymbol{\alpha}_{t}, \!\boldsymbol{\beta}_{t}) {=} \!\! \scaleobj{0.8}{\sum\nolimits_{\forall f^{\sf{m}} \in \mathcal{F}^{\sf{m}}}} \! R_{n,k}^{{\sf{m}},f^{\sf{m}},t_{s}^{\sf{m}}}(\boldsymbol{p}_{t}, \!\boldsymbol{P}_{t}, \!\boldsymbol{\alpha}_{t}, \!\boldsymbol{\beta}_{t}). \subnum
\eieq
Subsequently, assuming that ${\sf{UE}}_{k}^{\sf{m}}$ is served by the LSat over ${\sf{RB}}(f^{\sf{m}},t_{s}^{\sf{m}})$ of ${\sf{BWP}}_{t}^{\sf{m}}$, the corresponding SINR is given as
\beq
    \gamma_{0,k}^{f^{\sf{m}},t_{s}^{\sf{m}}}(\boldsymbol{p}_{t}, \boldsymbol{P}_{t}, \boldsymbol{\alpha}_{t}, \boldsymbol{\beta}_{t}) = \scaleobj{0.8}{\frac{\beta_{k}^{f^{\sf{m}},t_{s}^{\sf{m}}} P_{k}^{f^{\sf{m}},t_{s}^{\sf{m}}} g_{k}^{f^{\sf{m}},t_{s}^{\sf{m}}}}{ \Theta_{k}^{{\sf{s}},f^{\sf{m}},t_{s}^{\sf{m}}}(\boldsymbol{p}_{t}, \boldsymbol{\alpha}_{t}) + \sigma_{{\sf{m}},k}^2}}, \nonumber
\eeq
where $\Theta_{k}^{{\sf{s}},f^{\sf{m}},t_{s}^{\sf{m}}}(\boldsymbol{p}_{t}, \boldsymbol{\alpha}_{t}) =  \sum_{\forall n} \sum_{\forall k'} \alpha_{n,k'}^{f^{\sf{m}},t_{s}^{\sf{m}}} p_{n,k'}^{f^{\sf{m}}, t_{s}^{\sf{m}}} h_{n,k}^{f^{\sf{m}},t_{s}^{\sf{m}}}$ is the ISI caused by APs to ${\sf{UE}}_{k}^{\sf{m}}$ served by SN systems.
Therefore, the corresponding achievable rate
of ${\sf{UE}}_{k}^{\sf{m}}$ 
can be given as
\bieq{ll} \label{eq: rate RB ms sat}
    R_{0,k}^{{\sf{m}},f^{\sf{m}},t_{s}^{\sf{m}}} \! \!(\boldsymbol{p}_{t}, \! \boldsymbol{P}_{t},  \!\boldsymbol{\alpha}_{t}, \! \boldsymbol{\beta}_{t}) {=} w_{\sf{m}} \ln  \!  \left(  \!1 + \gamma_{k}^{f^{\sf{m}}, t_{s}^{\sf{m}}}(\boldsymbol{p}_{t}, \! \boldsymbol{P}_{t}, \! \boldsymbol{\alpha}_{t},  \! \boldsymbol{\beta}_{t}) \right), \subnum \\ 
    R_{0,k}^{{\sf{m}},t_{s}^{\sf{m}}}(\boldsymbol{p}_{t},  \!\boldsymbol{P}_{t}, \! \boldsymbol{\alpha}_{t},  \!\boldsymbol{\beta}_{t}) {=}  \!  \!\scaleobj{0.8}{\sum\nolimits_{\forall f^{\sf{m}} \in \mathcal{F}^{\sf{m}} }}  \! \!R_{{\sf{s}},k}^{{\sf{m}},f^{\sf{m}},t_{s}^{\sf{m}}}(\boldsymbol{p}_{t}, \! \boldsymbol{P}_{t},  \!\boldsymbol{\alpha}_{t}, \! \boldsymbol{\beta}_{t}). \subnum 
\eieq
\enlargethispage{-0.05cm} 

\subsubsection{Transmission over BWP $\sf{s}$}
In this BWP, UEs using SS services are served by the LSat. Assuming that ${\sf{UE}}_{k}^{\sf{s}}$ is served by the LSat over ${\sf{RB}}(f^{\sf{s}},t_{s}^{\sf{s}})$ of ${\sf{BWP}}_{t}^{\sf{s}}$, 
its  SNR can be expressed as
$
    \gamma_{0,k}^{f^{\sf{s}},t_{s}^{\sf{s}}}(\boldsymbol{P}_{t}, \boldsymbol{\beta}_{t}) = {\beta_{k}^{f^{\sf{s}},t_{s}^{\sf{s}}} P_{k}^{f^{\sf{s}},t_{s}^{\sf{s}}} g_{k}^{f^{\sf{s}},t_{s}^{\sf{s}}}}/{ \sigma_{{\sf{s}},k}^2}. \nonumber
$
The corresponding achievable rate at ${\sf{UE}}_{k}^{\sf{s}}$ can be given as
$
    R_{0,k}^{{\sf{s}},f^{\sf{s}},t_{s}^{\sf{s}}}(\boldsymbol{P}_{t}, \boldsymbol{\beta}_{t}) = w_{\sf{s}} \ln\left( 1 + \gamma_{0,k}^{f^{\sf{s}}, t_{s}^{\sf{s}}}(\boldsymbol{P}_{t}, \boldsymbol{\beta}_{t}) \right). 
$

Therefore, the aggregated rate of ${\sf{UE}}_{k}^{\sf{s}}$ at TS $t_{s}^{\sf{s}}$ served by the LSat and TF $t$ can be computed as
\bieq{ll} \label{eq: rate ss sat}
    R_{0,k}^{{\sf{s}},t_{s}^{\sf{s}}}(\boldsymbol{P}_{t}, \boldsymbol{\beta}_{t}) = \scaleobj{0.8}{\sum\nolimits_{\forall f^{\sf{s}} \in \mathcal{F}^{\sf{s}}}} R_{0,k}^{{\sf{s}},f^{\sf{s}}, t_{s}^{\sf{s}}}(\boldsymbol{P}_{t}, \boldsymbol{\beta}_{t}), \subnum \\
    R_{0,k}^{{\sf{s}},t}(\boldsymbol{P}_{t}, \boldsymbol{\beta}_{t}) = \scaleobj{0.8}{\sum\nolimits_{\substack{\forall f^{\sf{s}} \in \mathcal{F}^{\sf{s}} \\ \forall t_{s}^{\sf{s}} \in \mathcal{T}_{t}^{\sf{s}}}}} R_{0,k}^{{\sf{s}},f^{\sf{s}}, t_{s}^{\sf{s}}}(\boldsymbol{P}_{t}, \boldsymbol{\beta}_{t}). \subnum
\eieq
\enlargethispage{-0.05 cm} 
To ensure the total transmit power at APs and the LSat not exceeding the transmit power budget, it must satisfy
\vspace{-2mm}
\bieq{ll}
    (C11)\!:\; \scaleobj{0.8}{\sum\nolimits_{\forall k} \sum\nolimits_{\sf{x} \in \{d,m\}} \sum\nolimits_{\forall f^{\sf{x}} \in \mathcal{F}^{\sf{x}}}} p_{n,k}^{f^{\sf{x}},t_{s}^{\sf{x}}} \leq p_{n}^{\sf{max}}, \forall (n, t_{s}^{\sf{x}}), \nonumber \\
    (C12)\!: \; \scaleobj{0.8}{\sum\nolimits_{\forall k} \sum\nolimits_{\sf{x} \in \{m,s\}} \sum\nolimits_{\forall f^{\sf{x}} \in \mathcal{F}^{\sf{x}}} } P_{k}^{f^{\sf{x}},t_{s}^{\sf{x}}} \leq P^{\sf{max}}, \forall (m, t_{s}^{\sf{x}}). \nonumber
\eieq

\subsubsection{Traffic and Queuing Model}
We assume that there are $K$ traffic flows corresponding to $K$ UEs, and the data arrival at each time instance of services $\sf{m}$ and $\sf{s}$ will be served at the next TF while that of service $\sf{d}$ will be served at the next SF. 
Furthermore, one assumes that the arrival data of DS, MS, and SS services follows the Poisson distribution with means $\bar{\lambda}^{\sf{d}}$, $\bar{\lambda}^{\sf{m}}$, and $\bar{\lambda}^{\sf{s}}$, respectively. 
Besides, the DS and SS traffic flows at the core network (CN) are routed to the TN and SN systems, respectively, while the MS traffic flows are steered from the CN to TN and SN systems based on the traffic steering scheduler. 
We assume that the traffic steering decision is made for each TC.
First, we introduce the flow-spit decision variable for the MS traffic at the core network $\boldsymbol{\omega}_{c}^{\sf{cn}} = [ \omega_{k,c}^{\sf{cn}} ], \forall k \in \mathcal{U}^{\sf{m}}$, wherein $\omega_{k,c}^{\sf{cn}} \in [0,1]$ and $(1 - \omega_{k,c}^{\sf{cn}})$ are the portions of ${\sf{UE}}_{k}^{\sf{m}}$'s traffic which are steered to TN and SN systems during TC $c$, respectively.
Besides, TNs have multiple APs while SNs have one LSat. Hence, the traffic flows at the TN CU are steered to APs. Herein, we introduce the flow-spit decision variables in TNs $\boldsymbol{\omega}_{c}^{\sf{x}} = [\omega_{n,k,c}^{\sf{x}}], \forall (n,k) \in (\mathcal{N} \times \mathcal{U}^{\sf{x}}), {\sf{x}} \in \mathcal{S}^{\sf{tn}}$ wherein  $\omega_{n,k,c}^{\sf{x}} \in [0,1]$ indicates the flow split portion of service $\sf{x}\in \mathcal{S}^{\sf{tn}}$ from the TN CU to $\sf{AP}_{n}$ at TC $c$. Besides, the integrity of traffic flows is ensured by constraints
\vspace{-2mm}
\bieq{ll} \label{eq: flow-split}
    (C13)\!:  \scaleobj{0.8}{\sum\nolimits_{\forall n \in \mathcal{N} }} \omega_{n,k,c}^{\sf{x}} = 1, \forall k \in  \mathcal{U}^{\sf{x}}, {\sf{x}} \in \{{\sf{d}, \sf{m} } \}, \forall c. \nonumber 
\eieq

Subsequently, the data arrival of the DS, MS, and DS services of UEs at ${\sf{AP}}_{n}$ and the LSat are expressed as
\vspace{-2mm}
\bieq{ll} \label{eq: traffic arrival}
    \lambda_{n,k,t_{v}}^{\sf{d}} = \omega_{n,k,c}^{\sf{d}} \lambda_{k,t_{v}}^{\sf{d}}, \quad 
    \lambda_{n,k,t}^{\sf{m}} = \omega_{k,c}^{\sf{cn}} \omega_{n,k,c}^{\sf{m}} \lambda_{k,t}^{\sf{m}},  \subnum\\
    \lambda_{0,k,t}^{\sf{m}} = (1-\omega_{k,c}^{\sf{cn}})  \lambda_{k,t}^{\sf{m}}, \quad 
    \lambda_{0,k,t}^{\sf{s}} = \lambda_{k,t}^{\sf{s}}. \subnum 
\eieq

Regarding the DS services, their data packet size is usually small, however, these services has the strict delay requirement. Hence, we assume that the DS data arriving each time must be served within the next SF which yields the following constraint
\vspace{-2mm}
\beq
    (C14)\!: T_{\sf{d}} R_{n,k}^{f^{\sf{d}},t_{v}^{\sf{d}}} \! \geq \! \lambda_{n,k,t_{v}}^{\sf{d}}, \forall n, \forall k \in \mathcal{U}^{\sf{d}}, \forall t_{v}. \nonumber
\eeq
\enlargethispage{-0.05 cm} 


Assuming there is one buffer for each service, the queue length corresponding to traffic flow $k$ at ${\sf{BS}}_{n}$/${\sf{LSat}}$ of MS and SS services is expressed as
\bieq{ll}
    q_{n,k}^{{\sf{m}},t_{s}^{\sf{m}} + 1} = [q_{n,k}^{{\sf{m}},t_{s}^{\sf{m}}} +  \lambda_{n,k,t_{s}^{\sf{m}}}^{\sf{m}} - T_{\sf{m}} R_{n,k}^{{\sf{m}},t_{s}^{\sf{m}}} ]^{+}, \forall n, \forall k \in \mathcal{U}^{\sf{m}}, \nonumber \\
    q_{0,k}^{{\sf{x}},t_{s}^{\sf{x}} + 1} = [q_{0,k}^{{\sf{x}},t_{s}^{\sf{x}}} +  \lambda_{0,k,t_{s}^{\sf{x}}}^{\sf{x}} - T_{\sf{x}} R_{0,k}^{{\sf{x}},t_{s}^{\sf{x}}} ]^{+} ,\forall k \in \mathcal{U}^{\sf{x}}, \forall {\sf{x}} \in \mathcal{S}^{\sf{sn}}, \nonumber
\eieq
where $[\cdot]^{+} = \max \{0, \cdot\}$. Note that the data arrival of $\sf{x} \in \mathcal{S}^{\sf{sn}} $ will be served at the next TF. Hence, for $\sf{x} \in \mathcal{S}^{\sf{sn}} $, we set
\beq \label{eq: set data arrival}
\hspace{-2mm}
    \begin{cases}
        \lambda_{n,k,t_{s}^{\sf{x}}}^{\sf{x}} = \lambda_{n,k,t}^{\sf{x}}, \; \lambda_{0,k,t_{s}^{\sf{x}}}^{\sf{x}}=\lambda_{0,k,t}^{\sf{x}} \text{ if } s=1,  \\
        \lambda_{n,k,t_{s}^{\sf{x}}}^{\sf{x}}=0, \;\lambda_{0,k,t_{s}^{\sf{x}}}^{\sf{x}}= 0 \text{ in otherwise. }
    \end{cases}
\eeq
To ensure stability, the queue lengths must satisfy
\bieq{ll} \label{eq:queue-length}
    (C15)\!: q_{n}^{{\sf{m}},t_{s}^{\sf{m}}} = \scaleobj{0.8}{\sum\nolimits_{\forall k \in \mathcal{U}^{\sf{m}}} } q_{n,k}^{{\sf{m}},t_{s}^{\sf{m}}} \leq q_{n}^{\sf{m,max}}, \forall n, \forall t_{s}^{\sf{m}}, \nonumber \\
    (C16)\!: q_{0}^{{\sf{x}},t_{s}^{\sf{x}}} = \scaleobj{0.8}{\sum\nolimits_{\forall k \in \mathcal{U}^{\sf{x}}} } q_{0,k}^{{\sf{x}},t_{s}^{\sf{x}}} \leq q_{0}^{\sf{x,max}}, \forall t_{s}^{\sf{x}} , {\sf{x}} \in \mathcal{S}^{\sf{sn}}, \nonumber 
\eieq
wherein $q_{n}^{\sf{m,max}}$ and $q_{0}^{\sf{x,max}}$ are the maximum queue-length at ${\sf{AP}}_{n}$ and the LSat, respectively. For convenience, let's denote $\boldsymbol{q}_{c}^{\sf{tn}} = \{ q_{n}^{{\sf{m}},t_{s}^{\sf{m}}} | \forall n, \forall t_{s}^{\sf{m}} \in \mathcal{T}_{c}^{\sf{ts,m}} \}$, $\boldsymbol{q}_{c}^{\sf{sn}} = \{ q_{0}^{{\sf{x}},t_{s}^{\sf{x}}} | \forall t_{s}^{\sf{x}} \in \mathcal{T}_{c}^{\sf{ts,x}}, \forall {\sf{x}} \in \mathcal{S}^{\sf{sn}} \}$, and $\boldsymbol{q}_{c} \triangleq \{ \boldsymbol{q}_{c}^{\sf{tn}},\boldsymbol{q}_{c}^{\sf{sn}} \}$.

\begin{remark}
    The arrival DS data is served within the next SF ensured by $(C15)$. Hence, the corresponding buffer is cleared after each SF, the queue length of DS ones is not considered. 
\end{remark}
\enlargethispage{-0.05 cm} 
\subsection{Problem Formulation}
\subsubsection{Centralized Problem} 
Leveraging the predicted information provided by the DT  at the core-network in TC $c$,
we aim to minimize the system congestion by optimizing the BW allocation, traffic split decision, AP-UE and LSat-UE associations, RB assignment, and power control under the latency requirement of DS services. 
Hence, the objective function is the sum queue length of MS and SS services at APs and LSats in each TC $c$ which is defined as
\beq \label{eq: objective}
    f_{c}^{\sf{obj}}(\boldsymbol{q}_{c}) = \scaleobj{0.8}{\sum\nolimits_{\forall n, \forall t_{s}^{\sf{m}} \in \mathcal{T}_{c}^{\sf{ts,m}}} }   q_{n}^{{\sf{m}},t_{s}^{\sf{m}}} + \scaleobj{0.8}{\sum\nolimits_{\sf{x \in \mathcal{S}^{\sf{sn}}}, \forall t_{s}^{\sf{x}} \in \mathcal{T}_{c}^{\sf{ts,x}}} }  q_{0}^{{\sf{x}},t_{s}^{\sf{x}}}.
\eeq
The optimization problem is mathematically formulated as
\bieq{rll} \label{eq: Prob0}
    (\mathcal{P}_{0})_{c}: \; \min_{\boldsymbol{\Omega}_{c} } & \; & f_{c}^{\sf{obj}}(\boldsymbol{q}_{c}) \quad \st \; (C1)-(C16), \nonumber \\
    (C0): && b_{f^{\sf{x}},c}^{\sf{x}}, \alpha_{n,k}^{f^{\sf{x}},t_{s}^{\sf{x}}}, \beta_{m,k}^{f^{\sf{x}},t_{s}^{\sf{x}}} \in \{0,1\}, \forall (m,n,k,f^{\sf{x}},t_{s}^{\sf{x}},{\sf{x}},c), \nonumber
\eieq
where $\boldsymbol{\Omega_{c}} \triangleq
\{\boldsymbol{b}_{c},\boldsymbol{\omega}_{c},\boldsymbol{p}_{c}, \boldsymbol{P}_{c}, \boldsymbol{\alpha}_{c}, \boldsymbol{\beta}_{c} ,\boldsymbol{q}_{c}\}$.
Problem $(\mathcal{P}_{0})_{c}$ is very challenging to solve due to the coupling between the binary and continuous variables.

\subsubsection{Calibration Problem}
Since the prediction information of channel gain and arrival traffic may differ from that in the actual system, the  solution for problem $(\mathcal{P_{0}})$ should be adjusted appropriately. 
Specifically, 
based on actual channel gain at each TS, 
for the given $\boldsymbol{b}_{c}$, $\boldsymbol{\omega}_{c}$, $\boldsymbol{\alpha}_{c}, \boldsymbol{\beta}_{c}$, and $\boldsymbol{P}_{c}$,
the AP power calibration problem can be formulated as
\bieq{rll} \label{eq: Prob0Cali}
    (\mathcal{P}_{0}^{\sf{cali}})_{c}: \min_{\boldsymbol{p}_{c}, \boldsymbol{\eta}_{c}, \boldsymbol{q}_{c} } & & f_{c}^{\sf{obj}}(\boldsymbol{q}_{c}) 
    \quad \st \; (C10),(C11),(C14)-(C16). \nonumber
\eieq

\subsubsection{Overall work flow}
The work flow is summarized as\\
\textcircled{\small 1}Using update information, DT model predicts $\{\hat{\boldsymbol{h}}_{c}, \hat{\boldsymbol{g}}_{c}, \hat{\boldsymbol{\lambda}}_{c} \}$.\\
\textcircled{\small 2}Using $\{\hat{\boldsymbol{h}}_{c}, \hat{\boldsymbol{g}}_{c}, \hat{\boldsymbol{\lambda}}_{c}\}$, problem $(\mathcal{P}_{0})_{c}$ is solved for TC $c$.\\
\textcircled{\small 3}Based on actual $\{\boldsymbol{h}_{c}, \boldsymbol{g}_{c}, \boldsymbol{\lambda}_{c}\}$, and the given solution in step 2, the power control at APs is calibrated by solving $(\mathcal{P}_{0}^{\sf{cali}})$.

\enlargethispage{-0.05 cm} 
\section{Proposed Solutions}
\subsection{Compressed-Sensing-Based and Binary Relaxation}
To solve problem $(\mathcal{P}_{0})$ efficiently, the difficulty in solving due to binary variables should be addressed. 
Considering the system model and problem $(\mathcal{P}_{0})$, one can see the relationship between the variable pairs $\{ \boldsymbol{\alpha}_{t}, \boldsymbol{p}_{t} \}$ and $\{ \boldsymbol{\beta}_{t}, \boldsymbol{P}_{t} \}$ as follows.



\begin{itemize}
    \item If ${\sf{UE}}_{k}$ is served by ${\sf{AP}}_{n}$ via RB$(f^{\sf{x}}, t_{s}^{\sf{x}})$, $\alpha_{n,k}^{f^{\sf{x}},t_{s}^{\sf{x}}}=1$ and $p_{n,k}^{f^{\sf{x}},t_{s}^{\sf{x}}} > 0$ , otherwise $\alpha_{n,k}^{f^{\sf{x}},t_{s}^{\sf{x}}}=0$ and $p_{n,k}^{f^{\sf{x}},t_{s}^{\sf{x}}}=0$.
    \item If ${\sf{UE}}_{k}$ is served by the LSat via RB$(f^{\sf{x}}, t_{s}^{\sf{x}})$, $\beta_{k}^{f^{\sf{x}},t_{s}^{\sf{x}}}=1$ and $P_{k}^{f^{\sf{x}},t_{s}^{\sf{x}}}>0$, otherwise $\beta_{k}^{f^{\sf{x}},t_{s}^{\sf{x}}}=0$ and $P_{k}^{f^{\sf{x}},t_{s}^{\sf{x}}}=0$.
\end{itemize}
Based on this relationship, the $\boldsymbol{\alpha}_{t}$ and $\boldsymbol{\beta}_{t}$ can be respectively represented by $\boldsymbol{p}_{t}$ and $\boldsymbol{P}_{t}$ as
\bieq{ll}\label{eq: reduce binary}
    \hspace{-6mm} \alpha_{n,k}^{f^{\sf{x}},t_{s}^{\sf{x}}} = \Vert p_{n,k}^{f^{\sf{x}},t_{s}^{\sf{x}}} \Vert_{0}, \;
    \beta_{k}^{f^{\sf{x}},t_{s}^{\sf{x}}} = \Vert P_{k}^{f^{\sf{x}},t_{s}^{\sf{x}}} \Vert_{0}, \forall (n,m,k,f^{\sf{x}},t_{s}^{\sf{x}}). 
\eieq

Considering $(C4)$ and $(C7)$,  $b_{f^{\sf{x}},c}^{\sf{x}}=1$ if  $\alpha_{n,k}^{f^{\sf{x}},t_{s}^{\sf{x}}} =1$ or $\beta_{k}^{f^{\sf{x}},t_{s}^{\sf{x}}} = 1$. Hence, combine with \eqref{eq: reduce binary}, $\boldsymbol{b}$ is represented as
\bieq{ll} \label{eq: BW by norm0}
    b_{f^{\sf{d}},c}^{\sf{d}} \!\! =\!\! \Vert \hspace{-5mm} \scaleobj{0.8}{\sum_{\quad \forall (n,k), \forall t_{s}^{\sf{d}} \in \mathcal{T}_{c}^{\sf{ts,d}}}} \hspace{-6mm} \alpha_{n,k}^{f^{\sf{d}},t_{s}^{\sf{d}}} \Vert_{0} = \Vert p_{c}^{{\sf{d}},f^{\sf{d}}} \Vert_{0}, \; \forall (f^{\sf{d}}, c), \subnum \\
    b_{f^{\sf{m}},c}^{\sf{m}} \!\!=\!\! \Vert \hspace{-9mm} \scaleobj{0.8}{\sum_{\hspace{8mm} \forall (n,k), \forall t_{s}^{\sf{m}} \in \mathcal{T}_{c}^{\sf{ts,d}} }} \hspace{-8mm} \alpha_{n,k}^{f^{\sf{m}},t_{s}^{\sf{m}}} \!\!\! +\!\! \hspace{-8mm} \scaleobj{0.8}{\sum_{\hspace{8mm}\forall k, \forall t_{s}^{\sf{m}} \in \mathcal{T}_{c}^{\sf{ts,d}}} } \hspace{-7mm}\beta_{k}^{f^{\sf{m}},t_{s}^{\sf{m}}} \!\! \Vert_{0} 
    = \Vert p_{c}^{{\sf{m}},f^{\sf{m}}} \!\!\!\!+\!\! P_{c}^{{\sf{m}},f^{\sf{m}}} \!\! \Vert_{0}, \; \forall (f^{\sf{m}}, c), \hspace{6mm} \subnum \\
    b_{f^{\sf{s}},c}^{\sf{s}} \!\!=\!\!  \Vert \hspace{-5mm} \scaleobj{0.8}{\sum_{\hspace{5mm} \forall k, \forall t_{s}^{\sf{s}} \in \mathcal{T}_{c}^{\sf{ts,s}}} } \hspace{-5mm} \beta_{k}^{f^{\sf{s}},t_{s}^{\sf{s}}} \Vert_{0} = \Vert P_{c}^{{\sf{s}},f^{\sf{s}}} \Vert_{0}, \; \forall (f^{\sf{s}}, c). \subnum
\eieq
where $p_{c}^{{\sf{x}},f^{\sf{x}}} \!=\! \scaleobj{0.9}{\sum_{\forall (n,k) \\\forall t_{s}^{\sf{x}} \in \mathcal{T}_{c}^{\sf{ts,d}} }} p_{n,k}^{f^{\sf{x}},t_{s}^{\sf{x}}} $ and $P_{c}^{{\sf{x}},f^{\sf{x}}} \!=\! \scaleobj{0.9}{\sum_{\forall k \\\forall t_{s}^{\sf{x}} \in \mathcal{T}_{c}^{\sf{ts,d}} }} P_{k}^{f^{\sf{x}},t_{s}^{\sf{x}}} $.
Thanks to \eqref{eq: BW by norm0}, constraints $(C4)$ and $(C7)$ can be omitted.

Utilize relationships \eqref{eq: reduce binary}, the binary components in rate and SINR functions can be replaced by the corresponding $\ell_{0}$-norm components. Moreover, the binary variables in the production components of association and transmit power variables can be omitted. 
Specifically, $R_{n,k}^{{\sf{d}}, t_{v}^{\sf{d}}}(\boldsymbol{p}_{t},\boldsymbol{\alpha}_{t})$ is reformulated as
\bieq{ll} \label{eq: rate DS 2}
    R_{n,k}^{{\sf{d}}, t_{v}^{\sf{d}}}(\boldsymbol{p}_{t}) = w_{\sf{d}} \hspace{-6mm} \scaleobj{0.8}{ \sum_{\forall (f^{\sf{d}},t_{s}^{\sf{d}}) \in (\mathcal{F}^{\sf{d}} \times \mathcal{T}_{t_{v}}^{\sf{sf,d}} )} } \hspace{-8mm} \ln{(1 + \gamma_{n,k}^{f^{\sf{d}},t_{s}^{\sf{d}}}(\boldsymbol{p}_{t})}) 
    - \chi_{\sf{d}} \scaleobj{0.7}{\sqrt{ \hspace{-11mm} \sum_{\quad \quad\quad \forall (f^{\sf{d}},t_{s}^{\sf{d}}) \in (\mathcal{F}^{\sf{d}} \times \mathcal{T}_{t_{v}}^{\sf{sf,d}} )} \hspace{-14mm}  \Vert p_{n,k}^{f^{\sf{d}},t_{s}^{\sf{d}}} \Vert_{0} }  }. \hspace{6mm}  
\eieq
Hence, arguments $\boldsymbol{\alpha}_{t}, \boldsymbol{\beta}_{t}$ in the rate/SINR functions is omitted.

Subsequently, to address the sparsity terms, the $\ell_{0}$-norm component $\Vert x \Vert_{0}, \forall x \geq 0$ can be approximated as $\Vert x \Vert_{0} \approx f_{\sf{ap}}(x)$ wherein $f_{\sf{ap}}(x)\triangleq 1 - e^{-x/ \epsilon}, \; 0<\epsilon \ll 1 $ is a concave function. Moreover, let $f_{\sf{ap}}^{(i)}(x)$ be an upper bound of $f_{\sf{ap}}(x)$ at feasible point $(x^{(i)})$, the $\ell_{0}$-norm components in \eqref{eq: reduce binary}, \eqref{eq: BW by norm0} can be approximated at iteration $i$ as 
\bieq{ll}\label{eq: apx norm0}
 \Vert \iota \Vert_{0} \approx f_{\sf{ap}}(\iota) \leq f_{\sf{ap}}^{(i)}(\iota), \iota \in \{ p_{n,k}^{f^{\sf{x}},t_{s}^{\sf{x}}}, P_{k}^{f^{\sf{x}},t_{s}^{\sf{x}}} , p_{c}^{f^{\sf{x}}}, P_{c}^{f^{\sf{x}}}  \},
\eieq
where upper bound $f_{\sf{ap}}^{(i)}(x)$ can be obtained based on \cite{Hung_VTC24} as
\beq \label{eq: fapx}
    f_{\sf{ap}}(x) \leq f_{\sf{ap}}^{(i)}(x) \triangleq 1/ \epsilon \exp(-x^{(i)} / \epsilon)(x - x^{(i)} - \epsilon) + 1.
\eeq
Applying approximations in \eqref{eq: reduce binary}, \eqref{eq: BW by norm0} and \eqref{eq: apx norm0}, by replacing binary component of $\boldsymbol{b}_{c}$, $\boldsymbol{\alpha}_{t}$, and $\boldsymbol{\beta}_{t}$ by the corresponding approximated $f_{\sf{ap}}^{(i)}(\cdot)$ in constraints $(C1)-(C3),(C5),(C6),(C8)-(C9)$, we directly obtained convex constraints which are named as $\boldsymbol{(\tilde{C}1)-(\tilde{C}3),(\tilde{C}5),(\tilde{C}6),(\tilde{C}8)-(\tilde{C}9)}$, respectively, and $(C10)$ is approximated by non-convex constraint $(\bar{C}10)$ as
\bieq{ll}
    (\bar{C}10): \; \gamma_{n,k}^{f^{\sf{d}},t_{s}^{\sf{d}}}(\boldsymbol{p}_{t}) \geq f_{\sf{ap}}^{(i)}(p_{n,k}^{f^{\sf{d}},t_{s}^{\sf{d}}}) \gamma_{0}^{\sf{d}}, \; \forall n, \forall k \in \mathcal{U}^{\sf{d}},\forall (f^{\sf{d}}, t_{s}^{\sf{d}}). \nonumber
\eieq

\begin{algorithm}[!t]
\scriptsize
\begin{algorithmic}[1]
    \captionsetup{font=scriptsize}
    \protect\caption{\
    \textsc{Predicted Information Algorithm with Re-optimization (PIAwRO)}}
    \label{alg: CenAlg}
    \STATE \textbf{Phase 1:} Optimization for TC $C$ with predicted $\hat{\boldsymbol{h}}_{c}$, $\hat{\boldsymbol{g}}_{c}$, $\hat{\boldsymbol{\lambda}}_{c}$ from DT.
    \STATE Set $i=1$ and generate an initial point $(\boldsymbol{p}_{c}^{(0)}, \boldsymbol{P}_{c}^{(0)}, \boldsymbol{\eta}_{c}^{(0)})$.\\
    \REPEAT
    \STATE Solve problem $(\mathcal{P}_{2})_{c}$ to obtain $(\boldsymbol{p}_{c}^{\star}, \boldsymbol{P}_{c}^{\star}, \boldsymbol{\eta}_{c}^{\star})$.
    \STATE Update $ (\boldsymbol{p}_{c}^{(i)}, \boldsymbol{P}_{c}^{(i)}, \boldsymbol{\eta}_{c}^{(i)})=(\boldsymbol{p}_{c}^{\star}, \boldsymbol{P}_{c}^{\star}, \boldsymbol{\eta}_{c}^{\star})$ and $i:=i+1$.
    \UNTIL Convergence
    \STATE Recovery binary solutions $\boldsymbol{\alpha}_{c}$, $\boldsymbol{\beta}_{c}$, and $\boldsymbol{b}_{c}$ by \eqref{eq: recover binary} and \eqref{eq: BW by norm0}.
    \STATE \textbf{Output 1:} The solution for problem $\{\boldsymbol{b}_{c}^{\star},\boldsymbol{\omega}_{c}^{\star},\boldsymbol{p}_{c}^{\star}, \boldsymbol{P}_{c}^{\star}, \boldsymbol{\alpha}_{c}^{\star}, \boldsymbol{\beta}_{c}^{\star} ,\boldsymbol{q}_{c}^{\star}\}$.
    \STATE \textbf{Phase 2:} Re-optimize with actual $\boldsymbol{h}_{t}, \! \boldsymbol{g}_{t},\!\boldsymbol{\lambda}_{t}$ and initial point by \textbf{Output 1}.
    \FOR{Each TF $t$ in TC $c$}
        \REPEAT
            \STATE Solve problem $(\mathcal{P}_{1}^{\sf{cali}})_{c}$ to obtain $(\boldsymbol{p}_{t}^{\star}, \boldsymbol{\eta}_{t}^{\star})$.
            \STATE Update $ (\boldsymbol{p}_{t}^{(i)}, \boldsymbol{\eta}_{t}^{(i)})=(\boldsymbol{p}_{t}^{\star}, \boldsymbol{\eta}_{t}^{\star})$ and $i:=i+1$.
        \UNTIL Convergence
    \ENDFOR
    \STATE \textbf{Output 2:} The adjusted AP power control  $\boldsymbol{p}_{c}^{\star}$.
\end{algorithmic}
\normalsize 
\end{algorithm}

\vspace{-2mm}
\enlargethispage{-0.05 cm} 
\subsection{Transform Traffic Steering and Queue Constraints}
To address non-convex constraints $(C14)-(C16)$,
we introduce an intermediate traffic split variable $\bar{\boldsymbol{\omega}} = \{ \bar{\omega}_{n,k,c}^{\sf{x}} | \forall (n,k,c,{\sf{x}}) \} $ and use it instead of $\omega_{k,c}^{\sf{cn}}$ and $\omega_{n,k,c}^{\sf{x}}$ with the relationship
$
 \bar{\omega}_{n,k,c}^{\sf{x}}= \omega_{n,k,c}^{\sf{d}} \text{ if } {\sf{x} \equiv d}, 
    \quad \bar{\omega}_{n,k,c}^{\sf{x}}=\omega_{k,c}^{\sf{cn}} \omega_{n,k,c}^{\sf{m}} \text{ if } {\sf{x} \equiv m}.
$
\enlargethispage{-0.05 cm} 
Hence, constraint $(C13)$ is rewritten  as
\bieq{ll} \label{eq: flow-split 1}
    (\tilde{C}13): \quad \scaleobj{0.8}{\sum\nolimits_{\forall n \in \mathcal{N} }} \bar{\omega}_{n,k,c}^{\sf{d}} = 1, \forall k \in  \mathcal{U}^{\sf{d}}, \forall c, \nonumber 
\eieq
Furthermore, the traffic arrival formula $\eqref{eq: traffic arrival}$ is rewritten as
\bieq{ll} \label{eq: traffic arrival 1}
    \lambda_{n,k,t_{v}}^{\sf{d}} = \bar{\omega}_{n,k,c}^{\sf{d}} \lambda_{k,t_{v}}^{\sf{d}}, \quad  
    \lambda_{n,k,t}^{\sf{m}} = \bar{\omega}_{n,k,c}^{\sf{m}} \lambda_{k,t}^{\sf{m}},  \subnum \label{eq: traffic arrival 1b}\\
    \lambda_{0,k,t}^{\sf{m}} = (1 - \scaleobj{0.8}{\sum\nolimits_{\forall n}} \bar{\omega}_{n,k,c}^{\sf{m}}) \lambda_{k,t}^{\sf{m}}, \quad 
    \lambda_{0,k,t}^{\sf{s}} = \lambda_{k,t}^{\sf{s}}. \subnum \label{eq: traffic arrival 1c}
\eieq

Additionally, we introduce slack variables $\bar{\boldsymbol{q}}_{c} \!=\! \{ \bar{q}_{n,k}^{{\sf{x}},t_{s}^{\sf{x}}} \!, \bar{q}_{0,k}^{{\sf{x}},t_{s}^{\sf{x}}} | \forall (n,k), \forall t_{s}^{\sf{x}} \! \in \! \mathcal{T}_{c}^{\sf{ts,x}}, \forall {\sf{x}} \! \in \! \mathcal{S} \}$ as the upper bound of the queue length and $\boldsymbol{r}_{c} \!=\! \{ r_{n,k}^{{\sf{x}},t_{s}^{\sf{x}}} \!, r_{n,k}^{{\sf{x}},f^{\sf{x}},t_{s}^{\sf{x}}} \!, r_{0,k}^{{\sf{x}},t_{s}^{\sf{x}}} \!, r_{0,k}^{{\sf{x}},f^{\sf{x}},t_{s}^{\sf{x}}} | \forall (n,k), \forall t_{s}^{\sf{x}} \! \in \! \mathcal{T}_{c}^{\sf{ts,x}}, \forall {\sf{x}} \! \in \! \mathcal{S} \}$ as the rate's lower bound, and relax $(C15)-(C16)$ as
\bieq{ll}
    (\tilde{C}14): \quad T_{\sf{d}} r_{n,k}^{f^{\sf{d}},t_{v}^{\sf{d}}} \! \geq \! \lambda_{n,k,t_{v}}^{\sf{d}}, \forall n, \forall k \in \mathcal{U}^{\sf{d}}, \forall t_{v}, \nonumber \\
    (\tilde{C}15a):  \bar{q}_{n,k}^{{\sf{m}},t_{s}^{\sf{m}}} + \lambda_{n,k,t_{s}^{\sf{m}}}^{\sf{m}} - T_{\sf{m}} r_{n,k}^{{\sf{m}},t_{s}^{\sf{m}}} \leq \bar{q}_{n,k}^{{\sf{m}},t_{s}^{\sf{m}} + 1}, \forall n, \forall t_{s}^{\sf{m}}, \nonumber \\
    (\tilde{C}15b): 0 \leq \bar{q}_{n,k}^{{\sf{m}},t_{s}^{\sf{m}}},  \scaleobj{0.8}{\sum\nolimits_{\forall k \in \mathcal{U}^{\sf{m}}}} \bar{q}_{n,k}^{{\sf{m}},t_{s}^{\sf{m}}} \leq q_{n}^{\sf{m,max}}, \forall n, \forall t_{s}^{\sf{m}}, \nonumber \\
    (\tilde{C}16a): \bar{q}_{0,k}^{{\sf{x}},t_{s}^{\sf{x}}} + \lambda_{0,k,t_{s}^{\sf{x}}}^{\sf{x}} - T_{\sf{x}} r_{0,k}^{{\sf{x}},t_{s}^{\sf{x}}} \leq \bar{q}_{0,k}^{{\sf{x}},t_{s}^{\sf{x}} + 1}, \forall t_{s}^{\sf{x}},\forall {\sf{x}} \in \mathcal{S}^{\sf{sn}}, \nonumber \\
    (\tilde{C}16b):  0 \leq \bar{q}_{0,k}^{{\sf{x}},t_{s}^{\sf{x}}},  \scaleobj{0.8}{\sum\nolimits_{\forall k \in \mathcal{U}^{\sf{x}}}} \bar{q}_{0,k}^{{\sf{x}},t_{s}^{\sf{x}}} \leq q_{0}^{\sf{x,max}}, \forall m, \forall t_{s}^{\sf{x}}, \forall {\sf{x}} \in \mathcal{S}^{\sf{sn}}, \nonumber 
\eieq
where $r_{n,k}^{{\sf{m}},t_{s}^{\sf{m}}} = \sum_{\forall f^{\sf{s}}} r_{n,k}^{{\sf{m}},f^{\sf{m}},t_{s}^{\sf{m}}}$, $r_{0,k}^{{\sf{x}},t_{s}^{\sf{x}}} = \sum_{\forall f^{\sf{s}}} r_{0,k}^{{\sf{x}},f^{\sf{x}},t_{s}^{\sf{x}}}, \forall {\sf{x}} \in \mathcal{S}^{\sf{sn}}$ while $r_{n,k}^{f^{\sf{x}},t_{s}^{\sf{x}}}$ and $r_{0,k}^{f^{\sf{x}},t_{s}^{\sf{x}}}$ satisfy constraints
\bieq{ll}
    (C17.1)\!: R_{n,k}^{{\sf{m}},f^{\sf{m}},t_{s}^{\sf{m}}}(\boldsymbol{p}_{t}) \geq r_{n,k}^{{\sf{m}},f^{\sf{m}},t_{s}^{\sf{m}}}, \; \forall n, \forall k \in \mathcal{U}^{\sf{m}}, \forall t_{s}^{\sf{m}} \in \mathcal{T}_{c}^{\sf{ts,m}}, \nonumber \\
    (C17.2)\!: R_{n,k}^{{\sf{d}},t_{v}^{\sf{d}}}(\boldsymbol{p}_{t}) \geq r_{n,k}^{{\sf{d}},t_{v}^{\sf{d}}}, \; \forall n, \forall k \in \mathcal{U}^{\sf{d}}, \forall t_{v}^{\sf{d}}, \nonumber \\
    (C18)\!: R_{0,k}^{{\sf{x}},f^{\sf{x}},t_{s}^{\sf{x}}}(\boldsymbol{P}_{t}) \geq r_{0,k}^{{\sf{x}},f^{\sf{x}},t_{s}^{\sf{x}}}, \; \forall k \in \mathcal{U}^{\sf{x}}, \forall t_{s}^{\sf{x}} \in \mathcal{T}_{c}^{\sf{ts,x}}, \forall {\sf{x}} \in \mathcal{S}^{\sf{sn}}, \nonumber
\eieq

Therefore, problem $(\mathcal{P}_{0})$ can be rewritten as
\bieq{rll} \label{eq: Prob1}
    (\mathcal{P}_{1})_{c}: &\; \min_{\bar{\boldsymbol{\omega}}_{c},\boldsymbol{p}_{c}, \boldsymbol{P}_{c}, \bar{\boldsymbol{q}}_{c}, \boldsymbol{r}_{c} } \quad f_{c}^{\sf{obj}}(\bar{\boldsymbol{q}}_{c}) \nonumber \\
    &\st \quad (\tilde{C}1)-(\tilde{C}3),(\tilde{C}5), (\tilde{C}6), (\tilde{C}8), (\tilde{C}9), (\bar{C}10), \nonumber \\
     & \hspace{8mm} (C11),(C12),(\tilde{C}14)-(\tilde{C}16), (C17),(C18). \nonumber
\eieq
Problem $(\mathcal{P}_{1})$ is still non-convex due to the non-convexity of $(\bar{C}10)$,$(C17)$,$(C18)$ convexified by the following Propositions.

\begin{figure}
    \centering
    \captionsetup{font=small}
    \includegraphics[width=5.5cm]{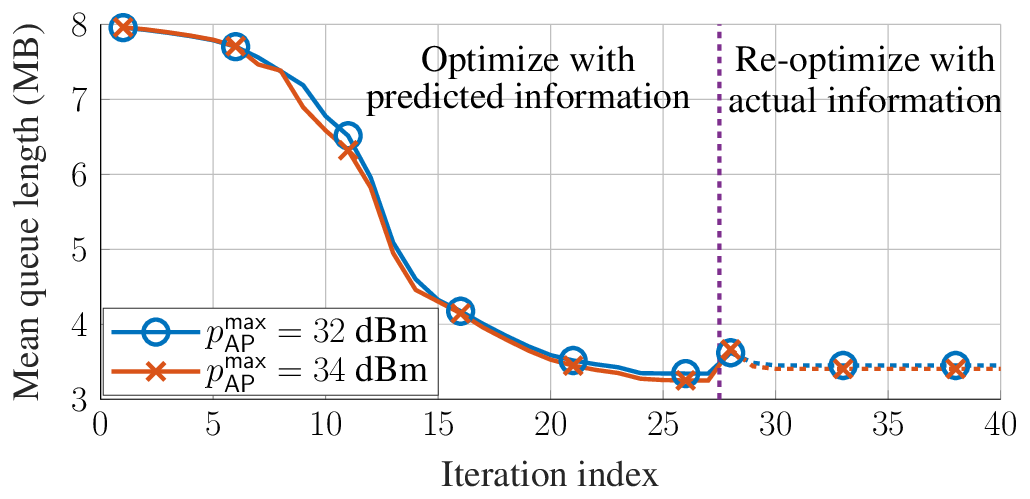}
    \vspace{-2mm}
    \caption{Convergence rate of the proposed algorithm.}
    \label{fig:Q_convergence}
    \vspace{-3mm}
\end{figure}
\enlargethispage{-0.05 cm} 
\begin{proposition} \label{pro: rate cons}
    Constraints $(C17.1),(C18)$ are convexified as
    \bieq{ll}
        (\tilde{C}17a)\!\!: f_{n,k}^{{\sf{rate}},f^{\sf{m}} \!\!,t_{s}^{\sf{m}}}(\boldsymbol{p}_{t}, \boldsymbol{P}_{t}, \boldsymbol{\eta}_{t}^{\sf{a}}) \geq r_{n,k}^{{\sf{m}},f^{\sf{m}},t_{s}^{\sf{m}}}, \nonumber \\
        (\tilde{C}17b)\!\!: \Psi_{n,k}^{f^{\sf{m}} \!\!,t_{s}^{\sf{m}}} \!(\! \boldsymbol{p}_{t} \!) \!+\!  \Theta_{k}^{{\sf{a}},\!f^{\sf{m}}\!\!,t_{s}^{\sf{m}}} \!\!(\!\boldsymbol{P}_{t}) + \sigma_{{\sf{m}},k}^{2} 
        \! \leq \! f_{\sf{exp}}^{(i)}(\eta_{n,k}^{{\sf{a}},f^{\sf{m}}\!\!,t_{s}^{\sf{m}}}) , \nonumber \\
        (\tilde{C}18a)\!\!: f_{0,k}^{{\sf{rate}},f^{\sf{m}} \!\!,t_{s}^{\sf{m}}}(\boldsymbol{p}_{t}, \boldsymbol{P}_{t}, \boldsymbol{\eta}_{t}^{\sf{s}}) \geq r_{0,k}^{{\sf{m}},f^{\sf{m}},t_{s}^{\sf{m}}}, \nonumber \\
        (\tilde{C}18b)\!\!: \Theta_{k}^{{\sf{s}},\!f^{\sf{m}}\!\!,t_{s}^{\sf{m}}} \!\!(\!\boldsymbol{p}_{t}) + \sigma_{{\sf{m}},k}^{2} 
        \! \leq \! f_{\sf{exp}}^{(i)}(\eta_{0,k}^{{\sf{s}},f^{\sf{m}}\!\!,t_{s}^{\sf{m}}}) , \nonumber \\
        (\tilde{C}18c)\!\!: R_{0,k}^{{\sf{s}},f^{\sf{s}},t_{s}^{\sf{s}}}(\boldsymbol{P}_{t}) \geq r_{0,k}^{{\sf{s}},f^{\sf{s}},t_{s}^{\sf{s}}}, \nonumber 
    \eieq
    with $f_{\sf{exp}}^{(i)}(u) \triangleq \exp(u^{(i)})(u - u^{(i)} + 1)$, $f_{n,k}^{{\sf{rate}},f^{\sf{m}} \!\!,t_{s}^{\sf{m}}}(\boldsymbol{p}_{t}, \boldsymbol{P}_{t}, \boldsymbol{\eta}_{t}^{\sf{a}}) = w_{\sf{m}} (\ln{\!( p_{n,k}^{f^{\sf{m}} \!\!,t_{s}^{\sf{m}}} h_{n,k}^{f^{\sf{m}} \!\!,t_{s}^{\sf{m}}} \!+\! \Psi_{n,k}^{f^{\sf{m}} \!\!,t_{s}^{\sf{m}}} \!(\! \boldsymbol{p}_{t} \!)} \!+\!  \Theta_{k}^{{\sf{a}},\!f^{\sf{m}}\!\!,t_{s}^{\sf{m}}} \!\!(\!\boldsymbol{P}_{t}) + \sigma_{{\sf{m}},k}^{2}) - \eta_{n,k}^{{\sf{a}},f^{\sf{m}}\!\!,t_{s}^{\sf{m}}})$, and $f_{0,k}^{{\sf{rate}},f^{\sf{m}} \!\!,t_{s}^{\sf{m}}}(\boldsymbol{p}_{t}, \boldsymbol{P}_{t}, \boldsymbol{\eta}_{t}^{\sf{s}}) = w_{\sf{m}} (\ln{\!( P_{k}^{f^{\sf{m}} \!\!,t_{s}^{\sf{m}}} g_{k}^{f^{\sf{m}} \!\!,t_{s}^{\sf{m}}}  \!+\!  \Theta_{k}^{{\sf{s}},\!f^{\sf{m}}\!\!,t_{s}^{\sf{m}}} \!\!(\!\boldsymbol{p}_{t}) + \sigma_{{\sf{m}},k}^{2})} - \eta_{0,k}^{{\sf{s}},f^{\sf{m}}\!\!,t_{s}^{\sf{m}}})$.
\end{proposition}

\begin{IEEEproof}
    Consider function $f_{\sf{r}}(x,y)=\ln(1+\frac{x}{y+a}), x,y \geq 0, a > 0$,
    constraint $f_{\sf{r}}(x,y) \geq z, \; z\geq 0$ is approximated as  \cite{Hung_VTC24}
    \vspace{-2mm}
    \begin{equation} \label{eq: apx rate} 
            \ln(x+y+a) \geq z + u, \quad
        y + a \leq f_{\sf{exp}}^{(i)}(u).
    \end{equation}
    Apply  $\eqref{eq: apx rate}$ to log component in rate functions 
    with corresponding values $x$, $y$, $a$, and $u$,
    $(C17.1)$ and $(C18)$ are convexified by $(\tilde{C}17)$ and $(\tilde{C}18)$, respectively.
\end{IEEEproof}


\begin{proposition} \label{pro: SINR cons}
Constraint $(\bar{C}10)$ can be approximated as
\bieq{ll}
    (\tilde{C}10a)\!\!: f_{n,k}^{{\sf{rate}},f^{\sf{d}} \!\!,t_{s}^{\sf{d}}}(\boldsymbol{p}_{t}, \boldsymbol{\eta}_{t}^{\sf{a}}) \geq f_{\sf{ap}}^{(i)}(p_{n,k}^{f^{\sf{d}},t_{s}^{\sf{d}}}) \ln(1+\gamma_{0}^{\sf{d}}) , \nonumber \\
    (\tilde{C}10b)\!\!: \Psi_{n,k}^{f^{\sf{d}} \!\!,t_{s}^{\sf{d}}} (\boldsymbol{p}_{t}) \!+\! \sigma_{{\sf{d}},k}^{2} 
    \! \leq \! f_{\sf{exp}}^{(i)}(\eta_{n,k}^{{\sf{a}},f^{\sf{d}}\!\!,t_{s}^{\sf{d}}}), \nonumber
\eieq
$f_{n,k}^{{\sf{rate}},f^{\sf{d}} \!\!,t_{s}^{\sf{d}}}(\boldsymbol{p}_{t}, \boldsymbol{\eta}_{t}^{\sf{a}}) = w_{\sf{d}} ( \ln{\!( p_{n,k}^{f^{\sf{d}} \!\!,t_{s}^{\sf{d}}} h_{n,k}^{f^{\sf{d}} \!\!,t_{s}^{\sf{d}}} \!+\! \Psi_{n,k}^{f^{\sf{d}} \!\!,t_{s}^{\sf{d}}} \!(\! \boldsymbol{p}_{t} \!)} \!+\! \sigma_{{\sf{d}},k}^{2})
    - \eta_{n,k}^{{\sf{a}},f^{\sf{d}}\!\!,t_{s}^{\sf{d}}})$
\end{proposition}
\begin{IEEEproof}
    Constraint $(\bar{C}10)$ is transformed equivalently as
    \begin{equation}
        \ln(1 +\gamma_{n,k}^{f^{\sf{d}},t_{s}^{\sf{d}}}(\boldsymbol{p}_{t}) )\geq f_{\sf{ap}}^{(i)}(p_{n,k}^{f^{\sf{d}},t_{s}^{\sf{d}}}) \ln(1+\gamma_{0}^{\sf{d}}), \; \forall n, \forall k,\forall (f^{\sf{d}}, t_{s}^{\sf{d}}). \nonumber
    \end{equation}
    This constraint is approximated similarly to Proposition~\ref{pro: rate cons}.
\end{IEEEproof}

\begin{proposition}\label{pro: rate DS}
Rate constraint $(C17.2)$ is approximated as
\bieq{ll}
    (\tilde{C}17c)\!\!: \scaleobj{0.8}{\sum_{\forall f^{\sf{d}} \in \mathcal{F}^{\sf{d}}}} \scaleobj{0.8}{\sum_{\forall t_{s}^{\sf{d}} \in \mathcal{T}_{t_{v}}^{\sf{sf,d}}}} f_{n,k}^{{\sf{rate}},f^{\sf{d}} \!\!,t_{s}^{\sf{d}}}(\boldsymbol{p}_{t}, \boldsymbol{\eta}_{t}^{\sf{a}})
    \!-\! \chi_{\sf{d}} f_{\sf{sqrt}}^{(i)}( \zeta_{n,k}^{t_{v}^{\sf{d}}}) \geq r_{n,k}^{{\sf{d}},t_{v}^{\sf{d}}} , \nonumber \\
    (\tilde{C}17d)\!\!: \zeta_{n,k}^{t_{v}^{\sf{d}}} \! \geq \! \scaleobj{0.8}{\sum\nolimits_{\forall f^{\sf{d}} \in \mathcal{F}^{\sf{d}}} \sum\nolimits_{\forall t_{s}^{\sf{d}} \in \mathcal{T}_{t_{v}}^{\sf{sf,d}}} } f_{\sf{ap}}^{(i)}(p_{n,k}^{f^{\sf{d}}\!\!,t_{s}^{\sf{d}}}), \nonumber 
\eieq
wherein $f_{\sf{sqrt}}^{(i)}(x) = 0.5{x}/{\sqrt{x^{(i)}}} + 0.5{\sqrt{x^{(i)}}}$.
\end{proposition}
\begin{IEEEproof}
The log component in $R_{n,k}^{{\sf{d}},t_{v}^{\sf{d}}}(\boldsymbol{p})$ is approximated as in Proposition~\ref{pro: rate cons}. Applying \eqref{eq: apx norm0} to $\ell_{0}$-norm terms, we obtained the upper bound of the summation in the second term of $R_{n,k}^{{\sf{d}},t_{v}^{\sf{d}}}(\boldsymbol{p})$ as $(\tilde{C}17d)$.
Using SCA technique to obtain upper bound $f_{\sf{sqrt}}^{(i)}(\zeta_{n,k}^{t_{v}^{\sf{d}}}) \geq f_{\sf{sqrt}}(\zeta_{n,k}^{t_{v}^{\sf{d}}})$, we complete the proof.
\end{IEEEproof}

Thanks to proposition~\ref{pro: rate cons}, \ref{pro: SINR cons} and \ref{pro: rate DS}, problem $(\mathcal{P}_{1})$ is transformed into an iterative convex problem $(\mathcal{P}_{2})$ as
\bieq{rll} \label{eq: Prob2}
    &&(\mathcal{P}_{2})_{c} : \min_{\boldsymbol{\omega}_{c},\boldsymbol{p}_{c}, \boldsymbol{P}_{c}, \bar{\boldsymbol{q}}_{c}, \boldsymbol{\eta}_{c}, \boldsymbol{\zeta}_{c} } \;f_{c}^{\sf{obj}}(\bar{\boldsymbol{q}}_{c}) \nonumber \\
    \st &&  (\!\tilde{C}1\!){-}(\!\tilde{C}3\!),(\!\tilde{C}5\!),(\!\tilde{C}6\!),(\!\tilde{C}8\!){-}(\!\tilde{C}10\!), 
     (\!C11\!),(\!C12\!),(\!\tilde{C}13\!){-}(\!\tilde{C}18\!). \nonumber
\eieq
After solving problem $(\mathcal{P}_{2})_{c}$, $\boldsymbol{\alpha}_{c}$ and $\boldsymbol{\beta}_{c}$ are recovered as
\bieq{ll} \label{eq: recover binary}
    \alpha_{n,k}^{f^{\sf{x}},t_{x}^{\sf{x}}} = 1 \text{ if } p_{n,k}^{f^{\sf{x}},t_{x}^{\sf{x}}} \geq \epsilon \text{ and } \alpha_{n,k}^{f^{\sf{x}},t_{x}^{\sf{x}}}=0 \text{ in otherwise, } \subnum \\
    \beta_{k}^{f^{\sf{x}},t_{x}^{\sf{x}}} = 1 \text{ if } P_{k}^{f^{\sf{x}},t_{x}^{\sf{x}}} \geq \epsilon \text{ and } \beta_{k}^{f^{\sf{x}},t_{x}^{\sf{x}}}=0 \text{ in otherwise. } \subnum
\eieq
The BW allocation variable $\boldsymbol{b}_{c}$ is recovered by \eqref{eq: BW by norm0}. Subsequently, using approximated convex constraints above, $(\mathcal{P}_{0}^{\sf{cali}})_{c}$ is transformed into the iterative convex problem $(\mathcal{P}_{1}^{\sf{cali}})_{c}$ as
\vspace{-2mm}
\bieq{rll} \label{eq: Prob1Cali}
    (\mathcal{P}_{1}^{\sf{cali}})_{c}:  \min_{\boldsymbol{p}_{c}, \boldsymbol{\eta}_{c}, \bar{\boldsymbol{q}}_{c} } &  & f_{c}^{\sf{obj}}(\bar{\boldsymbol{q}}_{c}) 
    \quad \st \; (\tilde{C}10),(C11),(\tilde{C}14)-(\tilde{C}16). \nonumber
\eieq
\vspace{-2mm}
The solution is summarized in Alg.~\ref{alg: CenAlg}, so-called PIAwRO.

\vspace{-1mm}
\section{Numerical Results}
\vspace{-2mm}
We set the operation frequency $f_c = 3.4$ GHz, system BW $B=15$ MHz, power budget at AP and LSat $p^{\sf{max}}_{\sf{ap}}=34$ dBm, $p^{\sf{max}}_{\sf{sat}}=36$ dBm, and number of DS, MS, SS UEs and APs $(K_{\sf{d}}, K_{\sf{m}}, K_{\sf{s}},N) = (4,5,3,6)$. The LEO altitude is $500$ km, and LSat and AP antenna parameters are set according to \cite{3gpp.38.811, 3gpp.38.863}. We consider $N_{\sf{cy}}=20$ TCs where each consists of $N^{\sf{tf}}=5$ TFs. For comparison purpose, we consider the following schemes:
1) FIA (Full information algorithm): use full actual information for optimization, 2) PIAwRO: proposed algorithm, 3) PIA: optimization with only predicted information (steps 1-8 of Alg.~\ref{alg: CenAlg}), and 4) Greedy algorithm: fixed BW allocation, $(\boldsymbol{\alpha}_{c},\boldsymbol{\beta}_{c})$ selected based on channel gain, water-filling-based power control, and traffic steering set proportional to UE rate.

Fig.~\ref{fig:Q_convergence} depicts the convergence rate in two phases of Alg.~\ref{alg: CenAlg} in different $p_{\sf{AP}}^{\sf{max}}$ values. One can see that Alg.~\ref{alg: CenAlg} converges after only about $30$ iterations. Especially, with the aid of DT, phase $1$ can provide a sufficiently good initial point so that phase $2$ only requires about $3$ iterations for convergence.

\begin{figure}
\vspace{-5mm}
    \centering
    \captionsetup{font=small}
    \includegraphics[width=8cm]{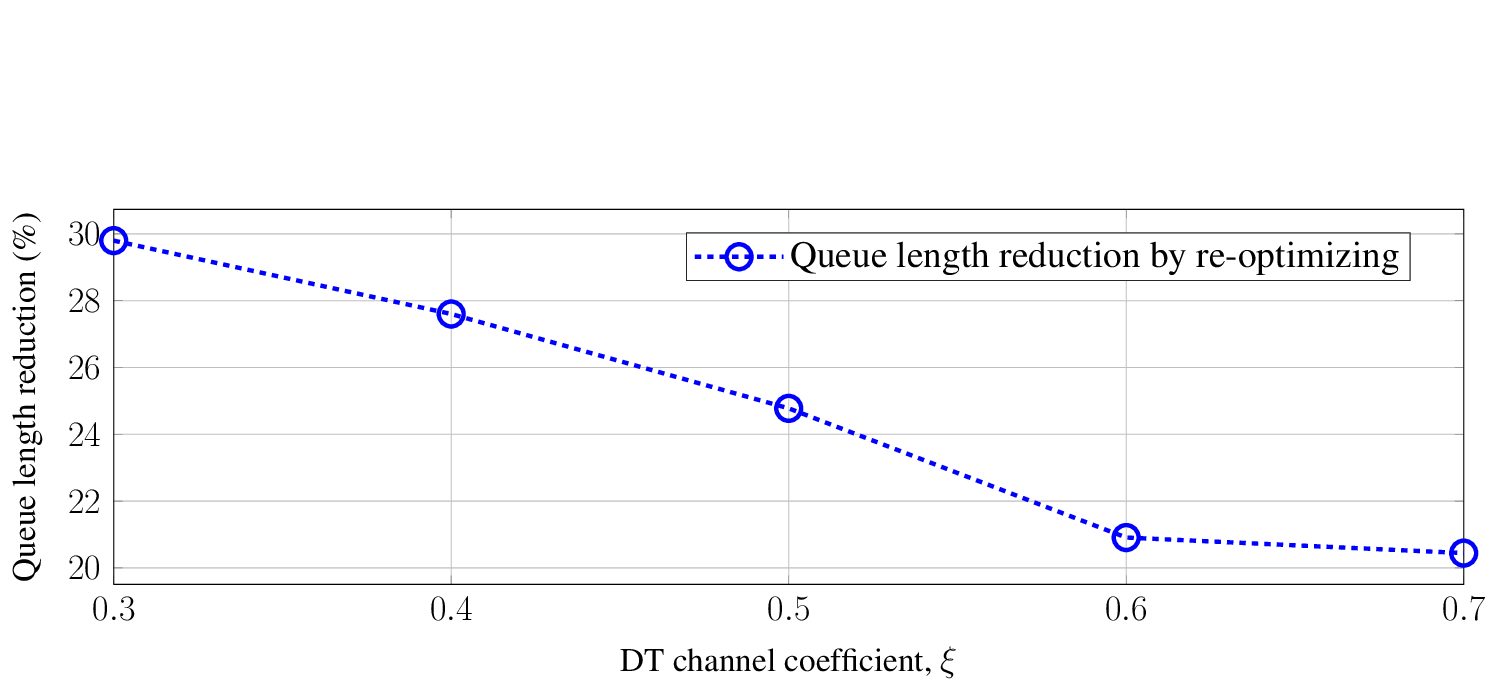}
    \vspace{-2mm}
    \caption{Mean queue length reduction vs. DT channel coefficient $\xi$.}
    \label{fig:Q_CoeDT}
    \vspace{-2mm}
\end{figure}

Fig.~\ref{fig:Q_CoeDT} shows the gain of re-optimizing in terms of queue length reduction versus the DT channel coefficient $\xi$. One notes that a larger $\xi$ indicates a more accurate replication of DT. Hence, the gain of re-optimizing decreases as $\xi$ increases due to the DT is closer to the actual environment.

\begin{figure}
    \vspace{-2mm}
    \centering
    \captionsetup{font=small}
    \includegraphics[width=7cm]{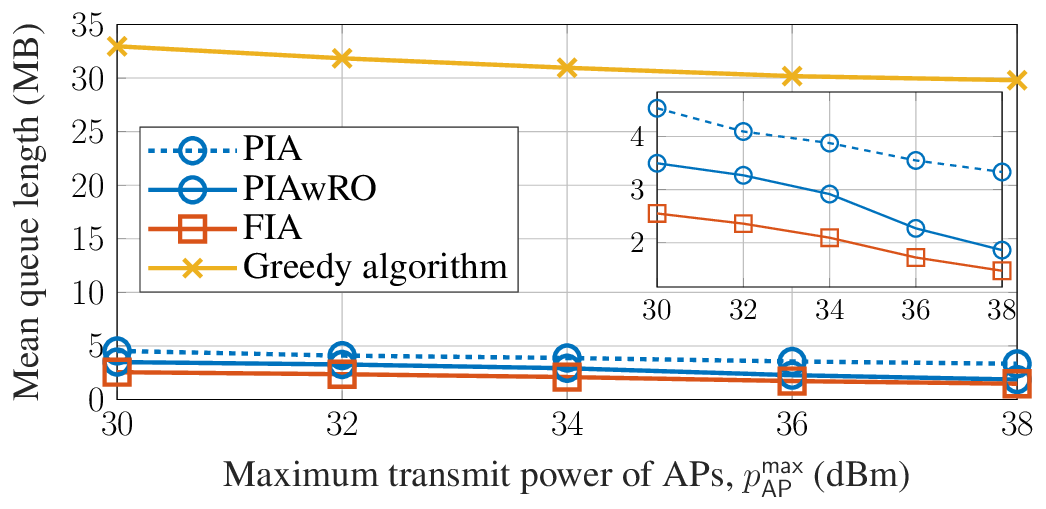}
    \vspace{-2mm}
    \caption{Mean queue length versus AP power budget $p_{\sf{AP}}^{\sf{max}}$.}
    \label{fig:Q_PA}
    \vspace{-2mm}
\end{figure}
Fig.~\ref{fig:Q_PA} shows the mean queue length outcome of the four considered schemes versus the AP power budget. Compared to the greedy algorithm, the three optimization-based algorithms are superior in minimizing the queue length. Especially, the gap between PIAwPO and FIA is only about $0.5-0.9$ MB. This figure further shows the gain of re-optimizing, i.e, re-optimizing $\boldsymbol{p}_{t}$ can reduce mean queue length by about $1$ MB.



\section{Conclusion}
\vspace{-1mm}
This work studied a DT-aided framework for resource management ISTNs, where TNs and SNs coexist over the same RFB. By leveraging a time-varying DT model with the 3D map, we jointly optimized BW allocation, traffic steering, UE/RB assignment, and power control, under delay service constraints. Subsequently, we proposed an SCA-based two-phase algorithm to solve the optimization problem. Numerical evaluations validated the superiority of our approach in minimizing queue lengths compared to benchmarks, highlighting its potential to support dynamic and spectrum-sharing ISTNs.

\vspace{-1mm}
\bibliographystyle{IEEEtran}
\bibliography{Journal}
\end{document}